\documentclass{emulateapj}

\usepackage{apjfonts}

\slugcomment{Draft}

\begin{document}

\shorttitle{Pop III Low-Mass Mode}
\shortauthors{A. Stacy and V. Bromm}

\title{The First Stars: A Low-Mass Formation Mode}

\author{Athena Stacy$^{1,2}$\thanks{E-mail: athena.stacy@berkeley.edu} and Volker Bromm$^{3}$}
\affil{$^{1}$NASA Goddard Space Flight Center, Greenbelt, MD 20771, USA \\
        $^{2}$University of California, Berkeley, CA 94720, USA \\
        $^{3}$Department of Astronomy and Texas Cosmology Center, University of Texas, Austin, TX 78712, USA}

\begin{abstract}
We perform numerical simulations of the growth of a Population III stellar system under photodissociating feedback.  We start from cosmological initial conditions at $z=100$, self-consistently following the formation of a minihalo at $z=15$ and the subsequent collapse of its central gas to high densities.  The simulations resolve scales as small as $\sim$ 1 AU, corresponding to gas densities of 10$^{16}$ cm$^{-3}$.  Using sink particles to represent the growing protostars, we evolve the stellar system for the next 5000 yr.  We find that this emerging stellar group accretes at an unusually low rate compared with minihalos which form at earlier times ($z=20-30$), or with lower baryonic angular momentum. The stars in this unusual system will likely reach masses ranging from $< 1$ M$_{\odot}$ to $\sim$ 5 M$_{\odot}$ by the end of their main-sequence lifetimes, placing them in the mass range for which stars will undergo an asymptotic giant branch (AGB) phase.
Based upon the simulation, we predict the rare existence of Population~III stars that have survived to the present day and have been enriched by mass overflow from a previous AGB companion. 
\end{abstract}


\section{Introduction}
Following the emission of the cosmic microwave background, the universe entered a period referred to as the `dark ages', when no luminous objects had yet formed. During this time, self-gravitating dark matter (DM) halos gradually grew in mass through the process of hierarchical merging. The dark ages ended when first stars, also known as Population III (Pop III), formed at $z \ga20$ within DM minihalos of mass $\sim$ 10$^6$ M$_{\odot}$ (e.g., \citealt{haimanetal1996,tegmarketal1997,yahs2003}).  

The typical mass of Pop III stars remains an open question that is crucial to our understanding of the evolution of the early universe (Bromm 2013). 
Their mass determines the rate at which they emitted ionizing radiation, and thus the extent to which they contributed to the reionization of the universe
(e.g., \citealt{kitayamaetal2004,syahs2004,whalenetal2004,alvarezetal2006,johnsongreif&bromm2007}).
 In addition, the Pop III mass determines how much they contributed to the metallicity of the intergalactic medium
 (IGM; \citealt{madauferrara&rees2001,moriferrara&madau2002,brommyoshida&hernquist2003,wada&venkatesan2003,normanetal2004,tfs2007,greifetal2007,greifetal2010,wise&abel2008,wise&abel2012,maioetal2011}; recently reviewed in \citealt{karlssonetal2013}).
For instance, Pop III stars with main sequence masses in the range 40~M$_{\odot}$~$<$~$M_{*}$~$<$~140~M$_{\odot}$ or  $M_{*}$~$>$~260~M$_{\odot}$ are expected to collapse directly into black holes, therefore releasing virtually no metals into the IGM.  On the other hand, stars with mass 140~M$_{\odot}$~$<$~$M_{*}$~$<$~260~M$_{\odot}$ are predicted to explode as pair-instability supernovae (PISNe; \citealt{heger&woosley2002}), thereby releasing the entirety of their metal content into the IGM and surrounding halos.  
We furthermore note recent work which has found that stellar rotation may significantly lower the PISN mass range (\citealt{chatz&wheeler2012, yoonetal2012}).  Primordial stars within the range  
15~M$_{\odot}$~$<$~$M_{*}$~$<$~40 M$_{\odot}$ will end their lives as core-collapse SNe, or possibly hypernovae in the case of rapid rotation (e.g., \citealt{nomotoetal2003}).  
Constraining the initial mass of Pop III stars is therefore central to understanding how the radiation and metallicity they released affected the formation of later stellar generations.

 Earlier work predicted that Pop III stars would form as single stars and grow to be very massive 
($\ga 100$ M$_{\odot}$; e.g., \citealt{abeletal2002,brommetal2002,bromm&loeb2004,yoh2008}).  
An analytical study by \cite{mckee&tan2008} found that even when accounting for radiative feedback, a protostar can grow to greater than 100~M$_{\odot}$.  Though a portion of the inflow towards the protostar will be reversed due to a growing ionization front (I-front), this I-front will expand preferentially in directions perpendicular to the protostellar disk, while accretion through the disk can continue unimpeded until much later times.

In contrast to the above picture of single and massive Pop III stars, more recent work has shown that primordial gas will undergo fragmentation and develop into a disk within which a stellar multiple system will form (e.g. \nocite{clarketal2008,clarketal2011a} Clark et al 2008; 2011a).  Such fragmentation is seen in simulations even when initialized on cosmological scales (e.g., \citealt{turketal2009,stacyetal2010}).  Furthermore, Pop III multiplicity occurs down to very small scales ($\sim$ 10 AU) and in the majority of minihalos  (Clark et al. 2011b, \nocite{clarketal2011b} \citealt{greifetal2011}), even when accounting for the effects of feedback from protostellar accretion luminosity (\citealt{smithetal2011, stacyetal2012}). 
Though a number of the above simulations exhibited disk fragmentation, the resolution study by \cite{turketal2012} found that increasing the number of resolution elements per Jeans mass leads to variation in gas morphology and suppression of disk formation.  They did not follow subsequent protostellar accretion, however, so whether fragmentation instabilities might later develop could not be determined.  \cite{latifetal2013} performed a similar study but followed the gas evolution beyond the formation of the first peak to find that self-gravitating disks with very rapid accretion rates  (10$^{-2}$ M$_{\odot}$ yr$^{-1}$) will indeed develop.

Several studies thus provide evidence that disk instability will develop after the first protostar arises in a primordial minihalo.  In addition, a fraction of minihalos may also be subject to 
earlier fragmentation in a pre-collapse phase.
%a chemothermal instability during the initial gas collapse.  
\cite{greifetal2013} find that this will lead to secondary clumps in at least two out of nine minihalos even before a protostar forms.
Though these recent studies generally imply a broader IMF for Pop III stars, the predicted IMFs remain top-heavy.  The primordial stellar systems still exhibit rapid accretion rates compared to modern-day star formation (e.g., \citealt{stacy&bromm2013}), and the most massive star in each system is still expected to eventually reach very  high masses ($\ga$ 10 M$_\odot$).  

In this paper we present the first three-dimensional numerical simulation to follow the growth of  a Pop III stellar system from cosmological initial conditions while also resolving nearly protostellar scales ($\sim$ 100 R$_{\odot}$) and accounting for the effects of photodissociating radiation.  Recent work has found that 100 R$_{\odot}$ ($\sim$ 1 AU) is approximately the maximum radius reached by a Pop III protostar during its pre-main sequence evolution (see e.g., \citealt{hosokawaetal2010}, \nocite{smithetal2011b} Smith et al 2011b).  Resolving these small scales corresponds to evolving the gas up to a maximum density of 10$^{16}$ cm$^{-3}$, at which point we continue the gas evolution for a further 5000 yr by employing sink particles to represent the growing protostars.
At this high density the gas is quickly approaching complete optical thickness to continuum radiation (10$^{18}$ cm$^{-3}$) and will not undergo further fragmentation on sub-sink scales (\citealt{yoh2008}).  
Our calculation therefore allows us to determine the true number of protostars that form within the central region of the host minihalo without missing any fragmentation due to lack of spatial resolution.  

While our numerical feedback model is also able to follow the effects of ionizing radiation from a growing protostar (e.g., \citealt{greifetal2009}), we find that the particular protostellar system we simulate does not contain stars sufficiently massive to produce an HII region.  This is an unusual system with significant variation from the more typical rapidly-accreting Pop III protostars studied in above-mentioned work.
%AGB STARS
Instead, the most massive star of the system considered here will most likely undergo an asymptotic giant branch (AGB) phase en route to becoming a white dwarf.

The AGB phase, which occurs for stars with initial masses between $\sim$ 0.8 and 8 M$_{\odot}$, plays an influential role in Galactic chemical evolution, so there is much interest in understanding the metal yields of low-metallicity AGB stars 
(e.g., \citealt{karakas2010,campbell&lattanzio2008,karakas2010, karakas&lugaro2010}). 
Similarly, if a significant population of AGB stars existed at high redshift, this would have consequences for metal and dust production in the early universe.  
Observations of high-redshift ($z\ga$ 6) galaxies and quasars indicate that significant amounts of dust had already formed at these early times, and the origin of such rapid dust production remains a subject of study (\citealt{bertoldietal2003,valianteetal2009,cherchneff&dwek2010,galletal2011}).   
Though SNe were likely the major source of dust at these early times, Pop III AGB stars could have provided a significant contribution as well. 
This applies in particular to stars such as the three most massive ones we present from our simulation, predicted to reach 3-5 M$_{\odot}$.  These stars have sufficiently short main-sequence lifetimes ($\sim$ 10$^8$ yr) to undergo an AGB phase by $z=6$.  In contrast, the smaller 1 M$_{\odot}$ stars from our simulation are too long lived to provide any dust or metallicity contribution.
The  more massive AGB stars could also have significantly contributed to carbon and nitrogen production in the early universe, as well as s-process elements (\citealt{bussoetal2001, siessetal2002, siess&goriely2003}).

Though much study of Pop III stars to date has emphasized the high-mass end of the Pop III IMF, we note that even some early studies 
predicted that typical Pop III stellar masses might be quite low, $\la$ 1 M$_{\odot}$. For instance, \citet{kashlinsky&rees1983} emphasized the importance of angular momentum in determining the mass of Pop III stars, predicting that rotational effects would cause the primordial gas clouds to collapse into a dense disk.  Only after the disk cooled to $\sim 1000$\,K through H$_2$ line emission could fragmentation occur. \citet{nakamura&umemura2001} predicted that fragmentation of primordial filaments would lead to a bimodal IMF, with a peak at $\sim$ 1 M$_{\odot}$ as well as at 100  M$_{\odot}$.

Here, we further explore the possible parameter space for Pop~III star formation by modeling with high accuracy the growth of an unusually low-mass primordial system. Such cases are expected to be rare, since most Pop III systems have been found to contain one or more high-mass stars that likely dominated the overall Pop III impact on the IGM and later stellar generations.  However, such low-mass stars as found in our simulation were potential survivors to the present-day, and may in principle be discovered within the Milky Way halo or nearby dwarf galaxies.  
In Section 2 we describe our numerical methodology, in Section 3 we discuss our protostellar evolution model, and in Section 4 we present our results. We discuss the impact of a global Lyman–Werner (LW) background in Section 5, and we conclude in Section 6.

\section{Numerical Methodology}

\subsection{Initial Setup}
We carry out our investigation using {\sc gadget-2,} a widely-tested three-dimensional N-body and SPH code (\citealt{springel2005}). 
We begin with a 200 kpc (comoving) box containing 128$^3$ SPH gas particles and the same number of DM particles.  The simulation is initialized at $z=100$.   Positions and velocities are assigned to the particles in accordance with a 
$\Lambda$CDM cosmology with $\Omega_{\Lambda}=0.7$, $\Omega_{\rm M}=0.3$, $\Omega_{\rm B}=0.04$, $\sigma_8=0.9$, and $h=0.7$.  The gas and DM evolution is followed  until the first minihalo forms and its central gas density reaches 
$10^4$ cm$^{-3}$. 

Once the site of the first minihalo is determined, the simulation is performed at higher resolution, starting again at $z=100$.  The increased resolution is attained using a hierarchical zoom-in procedure 
(e.g.  \citealt{navarro&white1994,tormenetal1997,gaoetal2005}) in which four nested refinement boxes are placed within the cosmological box,  centered on the site where the minihalo will eventually form.  Within each refinement level, each particle from the lower level is replaced with eight `child' particles, so that in the most refined region the parent particle is replaced by 4096 child particles. The four refinement levels have lengths of 40, 35, 30, and 20 $h^{-1}$ kpc (comoving), so that the most highly refined level encompasses all the mass that will later be incorporated into the minihalo.  The most refined SPH particles have mass $m_{\rm SPH} = 5 \times 10^{-3}$ M$_{\odot}$, and
the resolution mass of the refined simulation is 
$M_{\rm res}\simeq 1.5 N_{\rm neigh} m_{\rm SPH}\la  0.3 $M$_{\odot}$, where $N_{\rm neigh}\simeq 40$ is the typical number of particles in the SPH smoothing kernel (e.g., \citealt{bate&burkert1997}).

\subsection{Cut-Out Technique and Particle Splitting}
%The computational timesteps of the calculation become smaller as the gas density grows.  
%To increase the computational efficiency of the simulation,  
To increase computational efficiency,
once the gas has reached densities of $10^{12}$ cm$^{-3}$ we employ a `cut-out' technique in which all gas and DM beyond 3 pc from the densest gas particle is removed.  
%By this point in the evolution, 
At this point the central star-forming gas is gravitationally bound and under minimal influence from the mass of the outer minihalo and more distant regions of the cosmological box.  The total mass of the cut-out is 3500 M$_{\odot}$, and the minimum density is $\sim 10^2$ cm$^{-3}$ 
(see, e.g., \citealt{stacyetal2012} for further details).
%The gas at the cut-out edge has a free-fall time of $\sim 10^7$ yr and will undergo little evolution over the next 5000 yr followed in the simulation.  Our cut-out technique leads to the propagation of a rarefaction wave starting from the cut-out edge due to the vacuum boundary condition.  However, this will only travel a distance of $c_{\rm s} \, t$, where $c_{\rm s}$ is the gas soundspeed ($\sim$ 2 km s$^{-1}$), and the time $t$ is 5000 yr.  This corresponds to an insignificant distance of $\sim$ 10$^{-2}$ pc (2000 AU) from the cut-out edge, over two orders of magnitude smaller than the 3 pc box size.

At the same time that we cut out the central 3 pc of the cosmological box, we also further increase the particle resolution so that collapse to densities of 10$^{16}$ cm$^{-3}$ can be properly followed.  
We thus replace each SPH particle with 8 child particles, each of which is placed randomly within the smoothing kernel of the parent particle.  The mass of the parent particle is then evenly divided amongst the child particles.  Each of these particles inherits the same chemical abundances, velocity, and entropy as the parent particle (see, e.g., \citealt{bromm&loeb2003}, \nocite{clarketal2011b} Clark et al. 2011b).  This ensures conservation of mass, internal energy, and linear momentum.  After this process, each SPH particle in the new cut-out simulation has a mass of $m_{\rm sph}=6\times10^{-4}$ M$_{\odot}$, and the new resolution mass is $M_{\rm res} \simeq 0.03$ M$_{\odot}$.  
 This final $M_{\rm res}$ is close to the mass of the pressure-supported atomic core which develops once the opacity limit is reached  (\citealt{yoh2008}), defining the point at which the protostar has first formed.

\subsection{Chemistry, Heating, and Cooling}
We utilize the same chemistry and thermal network as described in detail by \cite{greifetal2009} and used in \cite{stacyetal2012}.   In short, the code follows the abundance evolution of  
H, H$^{+}$, H$^{-}$, H$_{2}$, H$_{2}^{+}$, He, He$^{+}$, He$^{++}$, and e$^{-}$, as well as the three deuterium species D, D$^{+}$, and HD.  
All relevant cooling mechanisms, including H$_2$ collisions with  H and He as well as other H$_2$ molecules, are included.  The thermal network also accounts for cooling through  
H$_2$ collisions with protons and electrons, H and He collisional excitation and ionization, recombination, bremsstrahlung, and inverse Compton scattering.  

Further H$_2$ processes must also be included to properly model gas evolution to high densities.  For instance, the chemistry and thermal network includes three-body H$_2$ formation and the concomitant H$_2$ formation heating, which become important at  $n \ga 10^8$ cm$^{-3}$.  Furthermore, when $n \ga 10^9$ cm$^{-3}$, cooling through H$_2$ ro-vibrational lines becomes less effective as these lines grow optically thick.  This is accounted for using an escape probability formalism together with the Sobolev approximation (see \citealt{yoshidaetal2006,greifetal2011} for further details).  
 Simple fitting formulae are also available for estimating optically thick H$_2$ rates (e.g., \citealt{ripamonti&abel2004}).   However, \cite{hirano&yoshida2013} find that fitting formulae can overestimate the cooling rate, such as in cases when the gas has a subtantial degree of rotation.  This can lead to significant differences in gas evolution, such as accelerated gravitational collapse, when using fitting formulae as opposed to the more accurate Sobolev method.

The most important new process utilized in the thermal network is H$_2$ collision-induced emission (CIE) cooling, which becomes significant when $n \ga 10^{14}$ cm$^{-3}$ (\citealt{frommhold1994}).  
As described in \cite{greifetal2011}, the reduction of the CIE cooling rate due to the effects of continuum opacity is handled through the following prescription (\citealt{ripamontietal2002,ripamonti&abel2004}):
\begin{equation}
\Lambda_{\rm CIE, thick}=\Lambda_{\rm CIE, thin}\,{\rm min}\left(\frac{1-e^{-\tau_{\rm CIE}}}{\tau_{\rm CIE}},1\right)\mbox{\ ,}
\end{equation}
where
\begin{equation}
\tau_{\rm CIE}=\left(\frac{n_{\rm H_2}}{7\times 10^{15}\,{\rm cm}^{-3}}\right)^{2.8}\mbox{\ ,}
\end{equation}

\noindent $\Lambda_{\rm CIE, thin}$ is the CIE cooling rate in the optically thin limit, and $\Lambda_{\rm CIE, thick}$ that in optically thick conditions.
\cite{hirano&yoshida2013} compared gas evolution when continuum opacity effects are calculated using the above fitting formula, and when they are instead estimated using 3D raytracing.  They find that when the fitting formula is used the gas collapses to $\sim 10^{17}$ cm$^{-3}$ only slightly faster, by $\sim$ 1 yr.  The differences in the thermal evolution are also minimal between the two methods.  We thus expect the fitting formula above to be sufficiently accurate.  
However, while \cite{hirano&yoshida2013} modeled runaway gas collapse, it is possible that the differences would be more substantial when considering longer-term evolution of a disk, and this will be further examined in future work.  

\subsection{Sink Particle Method}
When an SPH particle reaches a density of $n_{\rm max} = 10^{16}$ cm$^{-3}$, it is converted to a sink particle 
(e.g., \citealt{bateetal1995,brommetal2002,marteletal2006}; see also \citealt{stacyetal2012} for further details on the employed sink particle method.)
The sink then grows in mass by accreting surrounding particles within its accretion radius, which we set equal to the resolution length such that
 $r_{\rm acc} = L_{\rm res} \simeq 1.0$ AU.
 %, where

%\begin{equation}
%L_{\rm res}\simeq \left(\frac{M_{\rm res}}{\rho_{\rm max}}\right)^{1/3} \mbox{\ .}
%\end{equation}

\noindent The sink accretes a gas particle within $r_{\rm acc}$ as long as the particle is not rotationally supported against infall onto the sink.  This is determined by checking that the particle  satisfies  $j_{\rm SPH} < j_{\rm cent}$, 
where $j_{\rm SPH} = {\rm v}_{\rm rot} d$ is the specific angular momentum of the gas particle, $j_{\rm cent} = \sqrt{G M_{\rm sink} r_{\rm acc}}$ the level required for centrifugal support, and v$_{\rm rot}$ and $d$ are the
rotational velocity and distance of the particle relative to the sink.  Particles that satisfy these criteria are removed from the simulation, and their mass is added to that of the sink.  When the sink first forms it immediately accretes most of the particles within its smoothing length, so its initial mass is near the resolution mass of the simulation, 
$M_{\rm res} \simeq 3 \times 10^{-2}$ M$_{\odot}$.  
Its position and velocity are set to the average of that of the accreted particles.  

These same accretion criteria are additionally used to determine whether two sinks may be merged.  However, we note that modifications to the sink merging algorithm can significantly alter the sink accretion history (\citealt{greifetal2011}). 
%and this will be further studied in future work. 
Recent work by \cite{greifetal2012} has resolved sub-protostellar scales (0.05 R$_{\odot}$) of primordial star-forming gas, tracking the merger rate of protostars by  following their interactions without using sinks.  They find that approximately half of the secondary protostars formed will indeed migrate towards and merge with the initial protostar.  
%Check this after the sim has run further:
Our merging algorithm leads to a similar fraction of secondary sinks merging with the main sink.  After the first sink arises, six secondaries later form, but two of them eventually merge with the initial sink.  This roughly agrees with what \cite{greifetal2012} find to occur on sub-sink scales.  

%When the sink accretes a new particle, its position and velocity are updated to the mass-weighted average of the sink and the accreted gas particle or secondary sink.   The sink is held at a constant density of $10^{16}$ cm$^{-3}$ and temperature of 2000 K, the typical temperature for gas at this density.  The sink thus exerts a pressure on the surrounding particles, avoiding the formation of an artificial pressure vacuum around its accretion radius 
%(see \citealt{bateetal1995,brommetal2002,marteletal2006}).  

%Because sink particles are no longer evolved to higher densities, we avoid the problem of ever-decreasing computational timesteps as the density grows.  This allows the surrounding star-forming gas to be evolved for significantly longer periods of time, up to thousands of free-fall times.  The simulation can thus follow the further disk formation and fragmentation of the gas while still resolving the protostellar accretion rate on very small scales.  

\subsection{Ray-tracing Scheme}
Once the first sink particle forms, it represents a newly formed protostar and is used as the point source for modeling the effects of LW radiation emanating from the protostar.  We use the same scheme as described in \cite{stacyetal2012}. Briefly, our ray-tracing module generates a spherical grid, consisting of $\sim$ 10$^5$ rays and 200 radial bins, centered around the first sink.  The minimum radius is set equal to the distance between the sink and the nearest neighboring SPH particle, and the grid is updated each time the ray-tracing is performed.  Most particles within  $r_{\rm acc}$ from the sink are accreted, so the minimum radius is usually close to 1.0 AU.  The bins are logarithmically spaced from the minimum distance to 3 pc, the size of the cut-out region. Each SPH particle within a bin then contributes its density and chemical abundances, proportional to its density squared, to the average values assigned to the bin.

%H$_2$ dissociation by LW radiation (11.2 to 13.6 eV) is described by
%\begin{equation}
%k_{{\rm H}_{2}} =  1.1\times 10^{8}\,f_{\rm shield}\,F_{\rm LW}~{\rm s}^{-1} 
%\end{equation}
%\noindent  (\citealt{abeletal1997}), where $F_{\rm LW}$ denotes the radiation flux, in units of   
% erg s$^{-1}$ cm$^{-2}$ Hz$^{-1}$, at $h\overline{\nu}=12.87$eV,
%and  $f_{\rm shield}$ is the factor by which  H$_{2}$ self-shielding reduces the LW dissociation rate. This self-shielding factor depends upon the  H$_{2}$ column density $N_{\rm H_2}$.  With the above ray-tracing scheme, we determine $N_{\rm H_2}$ along each ray by summing the contribution from each bin.  We then use results from \cite{draine&bertoldi1996} to determine the value for $f_{\rm shield}$.  We note a recent update to their $f_{\rm shield}$ fitting formula (\citealt{wolcottetal2011,wolcott&haiman2011}), but do not expect this to significantly affect the results for our particular case.  Because of the unusually high column densities within the molecular disk ($N_{\rm H_2} \ga 10^{26}$ cm$^{-2}$), we calculate  $f_{\rm shield}$ using equation 37 from  \cite{draine&bertoldi1996}, which is more accurate  for large $N_{\rm H_2}$ than their simple power-law expression in their equation 36. 

Also as in Stacy et al. (2012), we then use the ray-tracing scheme
to determine the H$_2$ column density $N_{\rm H_2}$, 
and from this we determine the
shielding factor $f_{\rm shield}$ with the fitting formula from \cite{draine&bertoldi1996}. 
We note a recent update to the $f_{\rm shield}$ prescription by \cite{wolcottetal2011} and \cite{wolcott&haiman2011}, but do not expect this to significantly affect the results for our particular case. 
%but also see\citealt{wolcottetal2011,wolcott&haiman2011}).
We combine this with a protostellar evolution model (see Section 3) in which we assume a blackbody spectrum
 with an effective temperature $T_{\rm eff}$, as specified by that model. We then determine the approximate 
LW radiation flux $F_{\rm LW}$, in units of   
 erg s$^{-1}$ cm$^{-2}$ Hz$^{-1}$, at $h\overline{\nu}=12.87$eV  (\citealt{abeletal1997}).
This finally allows for a determination of the H$_2$ dissociation rate,
\begin{equation}
k_{{\rm H}_{2}} =  1.1\times 10^{8}\,f_{\rm shield}\,F_{\rm LW}~{\rm s}^{-1} \mbox{,}
\end{equation}
to be included in our chemical network.

\begin{figure*}
\includegraphics[width=.8\textwidth]{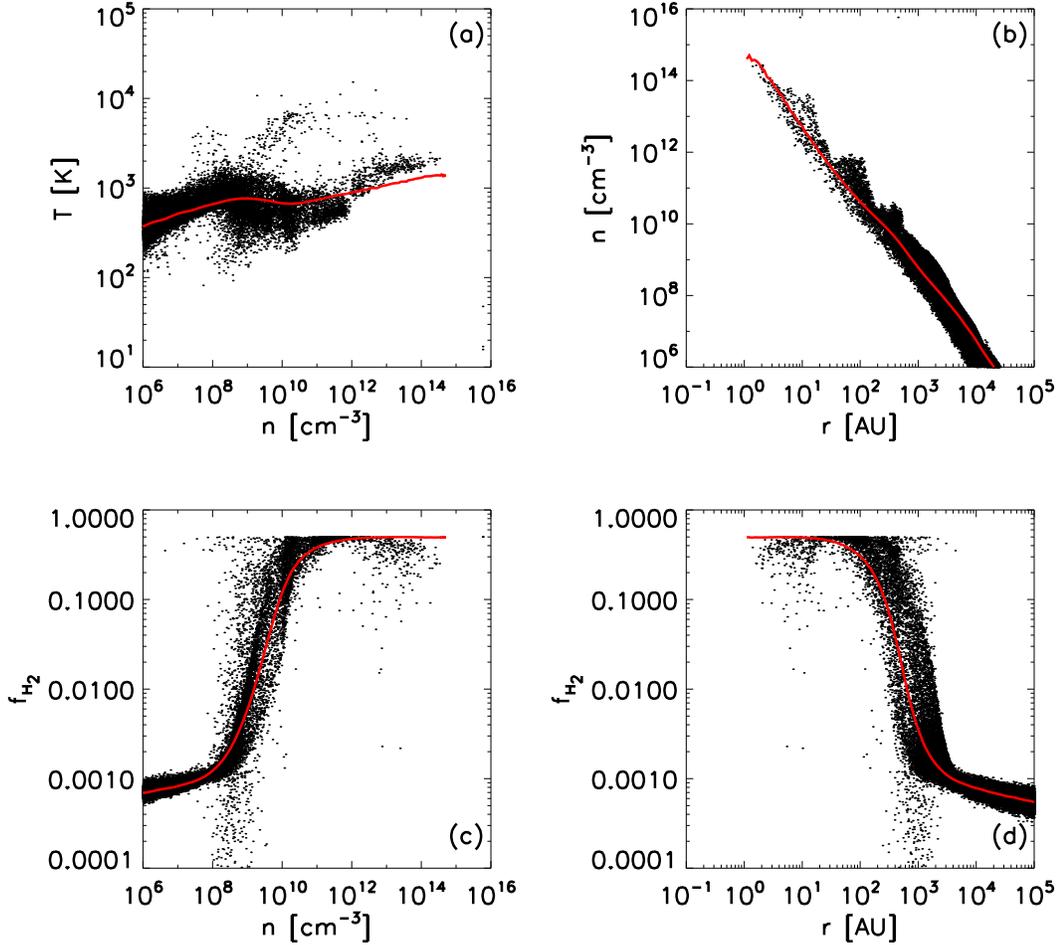}
 \caption{Physical state of the minihalo gas 3000 yr after the formation of the first sink particle.
{\it (a):}  Temperature versus number density $n$.
{\it (b):}  $n$ versus distance $r$ from the main sink particle.
{\it (c):} H$_2$ fraction ${f_{\rm H}}_2$ versus $n$.
%{\it (d):} Bonnor-Ebert mass $M_{\rm BE}$ vs. $n$.  Dashed line shows the resolution mass of our simulation, $3 \times 10^{-2}$ M$_{\odot}$
%{\it (d):} HD fraction versus  $n$.
{\it (d):} H$_2$ fraction versus distance $r$.
Red line shows the radially averaged values of the same quantities just prior to the formation of the first sink, as measured from the densest gas particle.
Note the warm phase of gas at $n \ga 10^7$ cm$^{-3}$, where the gas has been warmed through gravitational heating provided by the main sink.
The peak at 30 AU in panel {\it b} is where gas is accreting onto the secondary sink particle.  
}
\label{stuff-vs-nh}
\end{figure*}

\begin{figure}
\includegraphics[width=.45\textwidth]{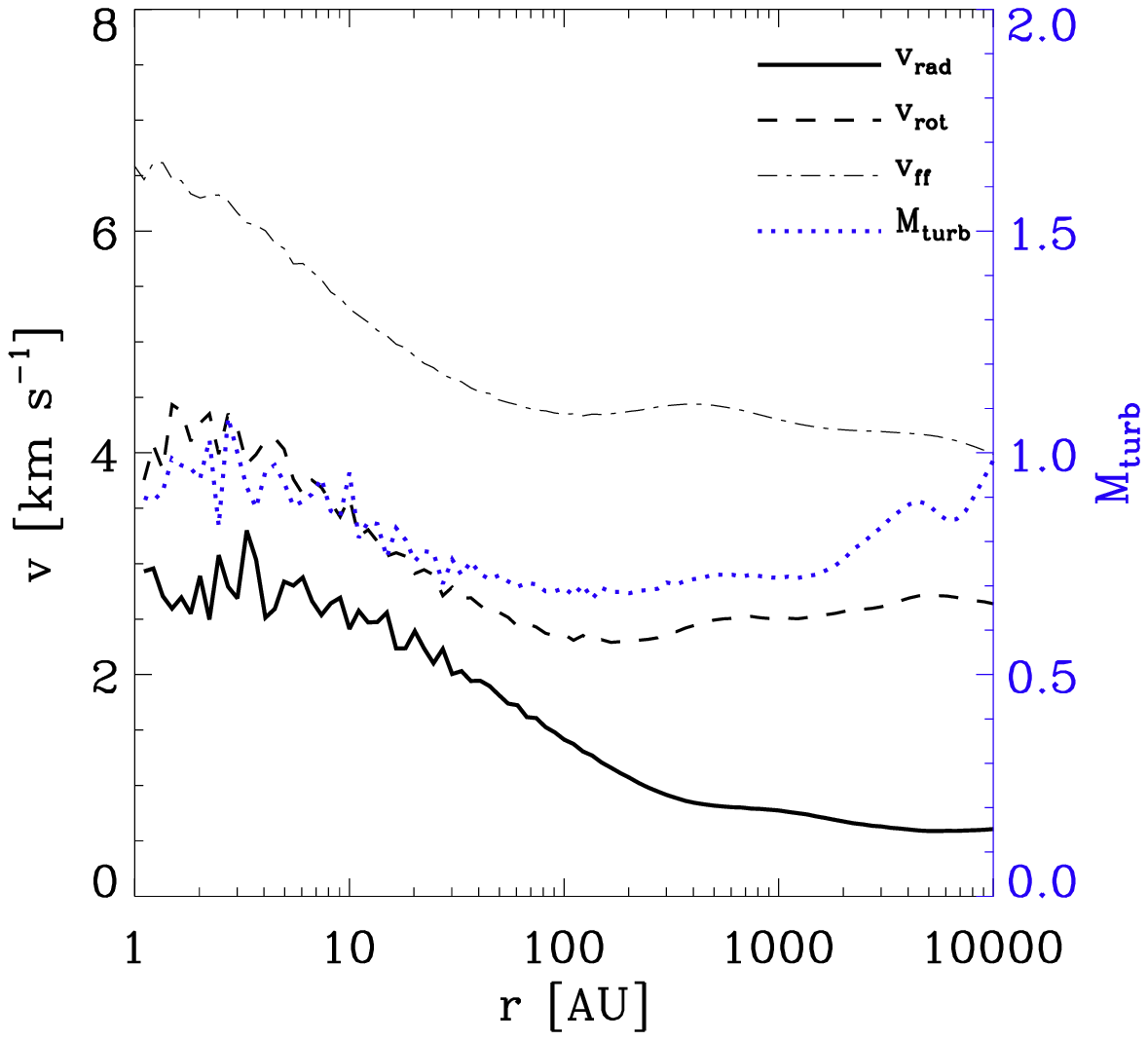}
\caption{
Velocity profile of the gas just prior to initial sink formation.
Solid line is the magnitude of the radial infall velocity $v_{\rm rad}$, averaged within a range of logarithmic radial bins.  Dashed line is the average rotational velocity $v_{\rm rot}$ within these same bins.  
Dash-dot line is the free-fall velocity $v_{\rm ff}$ based upon the enclosed mass at each radius. 
Dotted blue line is the turbulent Mach number $M_{\rm turb}$, which corresponds to the right-hand y-axis scale.  
The values of $v_{\rm rad}$ are low compared to $v_{\rm ff}$, and the magnitude of $v_{\rm rot}$ dominates over $v_{\rm rad}$ over all distances shown.
}
\label{velprof}
\end{figure}

\section{Protostellar Evolution Model}

\subsection{Luminosity and Temperature Evolution}

Our ray-tracing algorithm requires an input of protostellar effective temperature $T_{\rm eff}$ and luminosity $L_*$.  We calculate $L_*$ as the sum of $L_{\rm acc}$, the accretion luminosity,
and $L_{\rm int}$, the luminosity 
originating from the stellar interior and finally emitted from the photosphere of the protostar.  We write $L_*$ as

\begin{eqnarray}
L_* &=& L_{\rm acc} + L_{\rm int} 
= \alpha  \frac{G M_* \dot{ M}}{R_*} 
+  L_{\rm int} \mbox{,}
\end{eqnarray}

\noindent where $M_{*}$ is the protostellar mass, $\dot{M}$ the accretion rate,  and $R_*$ the protostellar radius (c.f. \citealt{prialnik&livio1985,hartmannetal1997}).  We take $M_*$ to be the mass of the sink, and $\dot{M}$ to be the accretion rate onto the sink, measured by averaging the total mass growth of the sink over the previous 
10 yr.  If the measured sink accretion rate yields $\dot{M} \simeq 0$, we simply assume the protostar is described by $L_* = L_{\rm int}$.

%The expression for $L_{\rm acc}$ is taken from, e.g., \cite{hartmannetal1997}, and 
We define $\alpha$ as the fraction of thermal energy from accretion that is added to the stellar interior.  Cold disk accretion would thus be described by $\alpha=0$, while for  hot spherical accretion $\alpha=1$.   For the main sink particle, we determine $\alpha$ by measuring $j_{\rm SPH}$ of each particle accreted by the sink within the last 10 yr,
as well as each particle currently within 10 AU from the sink.  
This allows us to find the percentage of nearby and recently accreted particles that have
%that were acquired with 
low angular momentum ($j_{\rm SPH} < 0.5j_{\rm cent}$) versus high angular momentum  ($j_{\rm SPH} > 0.5j_{\rm cent}$). 
As described in Section 2.5,  $j_{\rm SPH} = {\rm v}_{\rm rot} d$ is the angular momentum of the gas particle and $j_{\rm cent} = \sqrt{G M_{\rm sink} r_{\rm acc}}$ is the angular momentum required for centrifugal support against infall onto the sink. 
We take $\alpha$ to be the fraction of particles with $j_{\rm SPH} < 0.5j_{\rm cent}$, such that  $\alpha=1$ if the near-sink gas is dominated by radial instead of rotational motion.
%, but in the luminosity prescription we impose a minimum $\alpha$ of $1/7$.           

$L_{\rm int}$ will vary with the mass of the protostar.  When the protostar first forms and has low mass, we assume it is on the Hayashi track of the Hertzsprung-Russell diagram.  We approximate this by holding the protostar at an effective temperature of $T_{\rm Hay} = 4500$ K while its luminosity may vary.  This yields a `Hayashi track' luminosity of

\begin{eqnarray}
L_{\rm Hay} = 4 \pi R_*^2 \sigma_{\rm SB} T_{\rm Hay}^4 \mbox{.}
\end{eqnarray}

\noindent The precise value of $T_{\rm Hay}$ will vary from $\sim$ 3000 to 5000 K depending upon stellar mass and opacity.  Because the protostar we consider is metal-free, the opacity of the protostellar atmosphere will differ from that of a Pop I or Pop II star, and the resulting $T_{\rm Hay}$ will tend to be marginally higher for Pop III stars (e.g., \citealt{stahleretal1986}).  We thus choose  $T_{\rm Hay}$ to be in the upper end of this range and set it to 4500 K.
The uncertain value of initial $T_{\rm eff}$ (from 3000 to 5000 K) corresponds to a $L_{\rm Hay}$ that may vary by a factor of eight.  However, a $T_{\rm eff}$ even in the upper end of this range will not lead to significant feedback until after the protostar leaves the Hayashi track.  Our simulation results are thus not sensitive to the choice of initial $T_{\rm eff}$.

If the protostar grows sufficiently massive, we assume it eventually transitions to the Henyey track  (\citealt{henyeyetal1955}, see also \citealt{hansenetal2004}), 
and will gradually contract down to the main-sequence radius and commence hydrogen burning. However, the protostellar system in our simulation exhibits unusually low mass, and thus has much longer evolutionary timescales than the more common high-mass Pop III stellar systems.  We thus do not follow sufficiently long timescales for the protostars to reach these later stages, so we do not discuss them here.
Our protostellar model at these early times predicts a typical $L_*$ and $T_{\rm eff}$ of $\sim$ 100 L$_{\odot}$ and 4500 K.  This luminosity can roughly double during periods of rapid accretion due to the contribution from $L_{\rm acc}$, while the corresponding $T_{\rm eff}$ will increase by a few hundred kelvins.
At such low $T_{\rm eff}$, the fraction of luminosity in the LW band is negligible.  Combined with significant H$_2$ shielding in the disk, LW dissociation is unimportant at these early times.

\subsection{Radial Evolution}

As discussed in previous one-dimensional studies  (\citealt{stahler&palla1986,omukai&palla2003,hosokawaetal2010}), a growing protostar initially undergoes an `adiabatic accretion' phase, where $R_*$ grows with mass.  This continues approximately while  $t_{\rm acc} <  t_{\rm KH}$, where   

\begin{equation}
t_{\rm KH} = \frac{G M_*^2}{R_*L_*}  
\end{equation}

\noindent is the Kelvin-Helmholtz (KH) timescale and

\begin{equation}
t_{\rm acc} = \frac{M_*}{\dot{M}}
\end{equation}

\noindent is the accretion timescale.  The protostar will later begin KH contraction approximately when $t_{\rm acc} >  t_{\rm KH}$. 
 
\cite{hosokawaetal2010} modeled the evolution of a primordial protostars growing at $\dot{M} = 10^{-3} \rm M_{\odot} yr^{-1}$.
They found that, particularly during the `adiabatic' phase, accretion through a geometrically thin disk will lead to smaller protostellar radii than spherically symmetric accretion.  
The true accretion geometry is likely somewhere in between 
the idealized `disk' and `spherical' cases, with the infall beginning as nearly spherical and growing gradually more disk-like over time.  
%Below we describe the radial evolution of the protostar in the `spherical' case, in which the protostar gains the entirely of its mass through spherical accretion (Section 2.8.1).  
%We also describe the `spherical-to-disk' case, in which the protostar still initially grows through spherical accretion but later transitions to accretion through a disk (section 2.8.2).  Though the models we present describe the entire evolution of the protostar, we note that not all of these evolutionary phases are actually utilized in our simulation.  In particular, our simulation does not follow the protostellar evolution long enough to follow it's KH contraction to the MS.  However, we include descriptions of these phases for completeness.
Given the spherical accretion case described in \nocite{omukai&palla2003} Omukai \& Palla (2003, see also \citealt{stahler&palla1986}), the radial evolution during the adiabatic accretion phase can be described by the following expression:

\begin{equation}
R_{I,\rm sphere} \simeq 49 {\rm R_{\odot}} \left(\frac{M_*}{\rm M_{\odot}}\right)^{1/3} \left(\frac{\dot{M}}{\dot{M}_{\rm fid}}\right)^{1/3}   \mbox{\ .}
\end{equation}
      
\noindent  where  $\dot{M}_{\rm fid}\simeq 4.4\times 10^{-3}$~M$_{\odot}$\,yr$^{-1}$, the fiducial accretion rate used in the above-mentioned studies (see also \citealt{stacyetal2010}).

If a protostar transitions from spherical to disk accretion,
the radial evolution during the `adiabatic accretion' phase will be significantly different from the purely spherical case.  %Power-law modeling of `spherical' and `spherical-to-disk' accretion, taken directly from the study of primordial stars growing at $\dot{M} = 10^{-3} \rm M_{\odot} yr^{-1}$ in Hosokawa et al. (2010; their figure 11), is reproduced in Fig. \ref{star-rad}.  
%As in the model of \cite{hosokawaetal2010}, we set the transition in accretion geometry to occur for $M_* = 0.1$ M$_{\odot}$.  
%In our calculation we set the transition to occur when $\alpha < 0.2$.
For the range of accretion rates studied in \cite{hosokawaetal2010}, after transitioning to disk accretion, the radius rapidly declines due to the decrease in entropy brought to the stellar interior.  For pure disk accretion, this decline can be described by

%\begin{equation}
%R_{I,\rm disk} \simeq R_0 \left(\frac{M_*}{M_0}\right)^{-0.63 }  \mbox{\ ,}
%\end{equation}
\begin{equation}
R_{I,\rm disk} \simeq R_0 \left(\frac{M_*}{M_0}\right)^{-0.63 }  \mbox{\ ,}
\end{equation}

\noindent where $R_0$ and $M_0$ are the protostellar mass and radius at the point of transition from spherical to disk accretion.  
In our case, the protostar's accretion is disky even from initial sink formation, so we simply use the initial sink mass to set $M_0 = 0.045$ M$_{\odot}$.
Following the disk accretion model of Hosokawa et al. (2010; their figure 4), we approximate $R_0 = 1.7M_*^{(-1/3)} = 5$ R$_{\odot}$.

We subsequently set our radius in between the spherical and disk cases such that

\begin{equation}
R_{I} = \alpha R_{I,\rm sphere} + \left( 1 - \alpha \right)R_{I,\rm disk} \mbox{\ .}
\end{equation}

\noindent If the radial decline is unphysically rapid ($\dot{R} < -R/t_{\rm KH}$), we limit the rate of contraction to  be $\dot{R} = -R/t_{\rm KH}$.
The radial decline will continue until deuterium burning begins in the stellar interior, thus increasing the average entropy within the star, which occurs when $T_{\rm int} > 2 \times 10^6$ K.  
After this point the protostar again undergoes a roughly adiabatic expansion.
When a maximum radius is reached (see \citealt{hosokawaetal2010}), KH contraction to the main sequence will begin.  
However, our simulations do not follow the protostellar growth to these later stages because we are examining an atypical low-mass case in which the protostellar evolutionary timescales are very long compared to those of more massive protostars.

In our model $R_*$ expands to nearly 30 R$_{\odot}$ over the first few hundred years during a period of more spherical-type accretion.  
After a transition to more disk-type accretion, $R_*$ gradually declines as $\dot{R} = -R/t_{\rm KH}$ down to $\sim$ 15 R$_{\odot}$ by the end of the simulation.  

\begin{figure*}
\includegraphics[width=.4\textwidth]{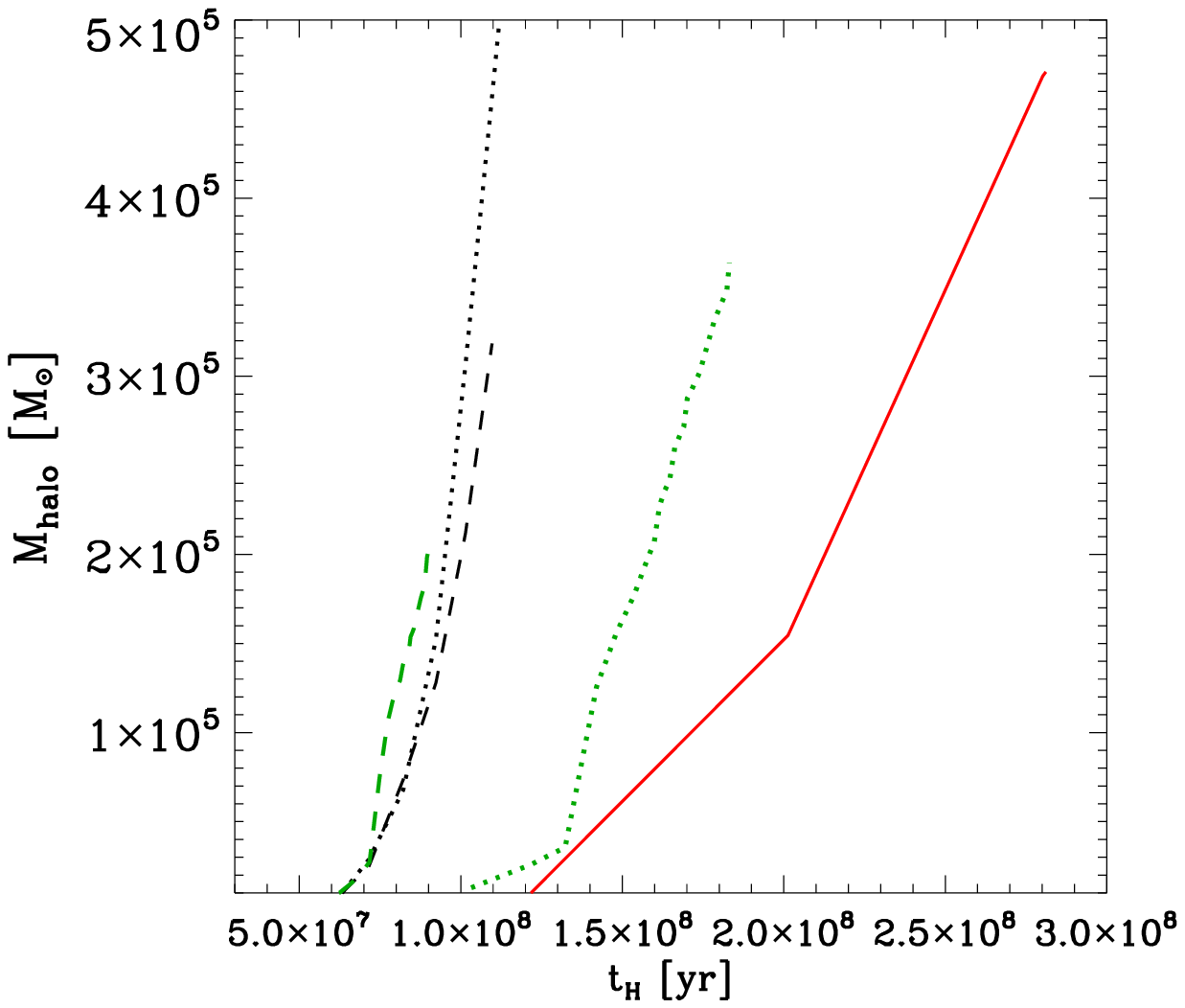}
\includegraphics[width=.4\textwidth]{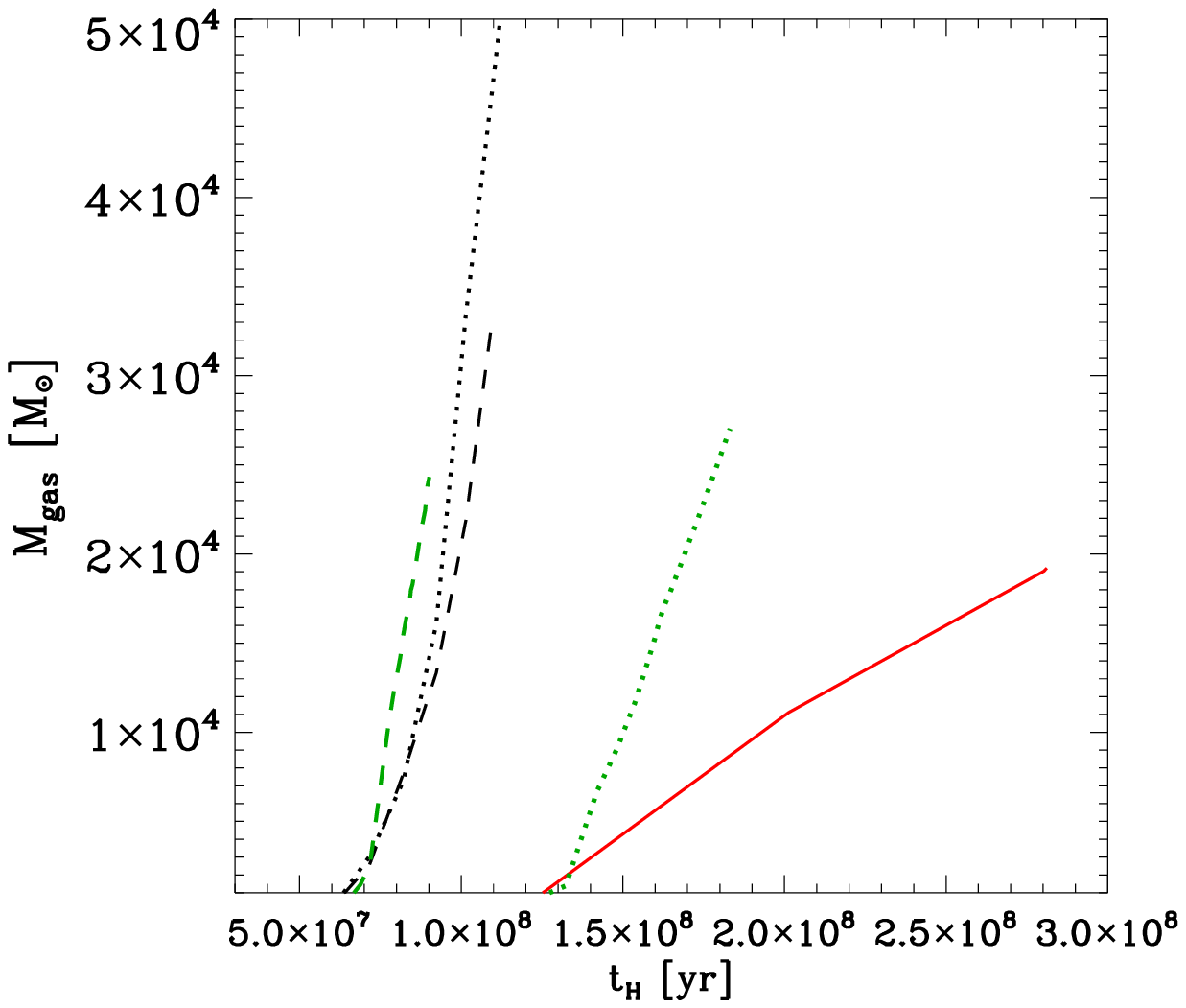}
\caption{
{\it Left:}  Evolution of the DM mass of various simulated halos over time, as measured in terms of age of the universe.
{\it Right:} Concurrent evolution of the baryonic mass within the halos.
The red line represents the $z=15$ minihalo of our simulation.
The black dotted line represents the Stacy \& Bromm (2013) minihalo which had the lowest overall stellar accretion rate, while the black dashed line represents that which had the greatest accretion rate.
Green dotted and dashed lines additionally show the growth of the two minihalos from Greif et al. (2012) which had the lowest and highest accretion rates, respectively.
%In both panels, dashed and dotted lines represent minihalos from simulations presented in Stacy and Bromm (2013) and Greif et al. (2012), as described in the previous figure.
}
\label{halomass}
\end{figure*}

\section{Results}

\subsection{Initial Minihalo Collapse}
The chemical and thermal evolution of the central minihalo gas up to the formation of the first sink particle is depicted by the red lines in  Figure \ref{stuff-vs-nh}.  The minihalo is in place by $z=15$, and the subsequent evolution is similar to that of previous work (e.g., \citealt{yoshidaetal2006,greifetal2011}).  The gas is heated through adiabatic compression as it approaches densities of 10$^8$ cm$^{-3}$. After this point, three-body reactions rapidly increase the H$_2$ fraction, such that the correspondingly enhanced cooling rate is similar to the combination of the H$_2$ formation heating rate and adiabatic heating rate due to collapse, yielding a roughly isothermal evolution.  The gas becomes fully molecular by densities of 10$^{10}$ cm$^{-3}$, forming  a $\sim 1$  M$_{\odot}$ molecular core.  Above these densities the H$_2$ cooling is no longer optically thin and the gas gradually heats again to $\sim$ 2000 K by $n = 10^{16}$ cm$^{-3}$, at which point the first sink particle forms.  

The velocity structure of the central 10,000 AU is shown in Figure \ref{velprof}.
Within each logarithmically-spaced
radial bin, we take $v_{\rm rad}$ and $v_{\rm rot}$  as the mass-weighted average of the individual particle velocities within each bin.    
Both the radial and rotational gas velocity $v_{\rm rad}$ and $v_{\rm rot}$ are on the order of half of $v_{\rm ff}$, where $v_{\rm ff}$ is the free-fall velocity based upon the enclosed mass at the given radius.  
The gas has a substantial amount of rotational support such that $v_{\rm rot}$  dominates over $v_{\rm rad}$ and is approximately half of the Keplerian velocity $v_{\rm Kep}$.

In a similar fashion,
we measure the turbulent Mach number, $M_{\rm turb}$, over the same range of 
radial bins according to:
\begin{equation}
M_{\rm turb}^2 c_s^2 =\sum_{i} \frac{m_i}{M}\left(\vec{\rm v}_i - \vec{\rm v}_{\rm rot }  - \vec{\rm v}_{\rm rad  }\right)^2 \mbox{,}
\end{equation}
where $c_s$ is the sound speed of the radial bin, $m_i$ is the mass of a gas particle contributing to the bin, and $M$ is the total gas mass of the bin.
The central 10,000 AU of gas are characterized by subsonic and nearly sonic turbulence.  

%\begin{table}
%\begin{tabular}[width=.45\textwidth]{crrrr}
%\hline
%sink  & $t_{\rm form}$ [yr] & $M_{\rm final}$ [M$_{\odot}$]  & $r_{\rm init}$ [AU] & $r_{\rm final}$ [AU]\\
%\hline
%1  & 0  & 19  & 0 & 0\\
%2  & 200 &  9.4 & 110 & 440\\
%\hline
%\end{tabular}
%\caption{Same as Table 1, but for sinks remaining at the end of the `with-feedback' case at 4500 yr.}
%\label{tab2}
%\end{table}

\begin{figure*}
\includegraphics[width=.4\textwidth]{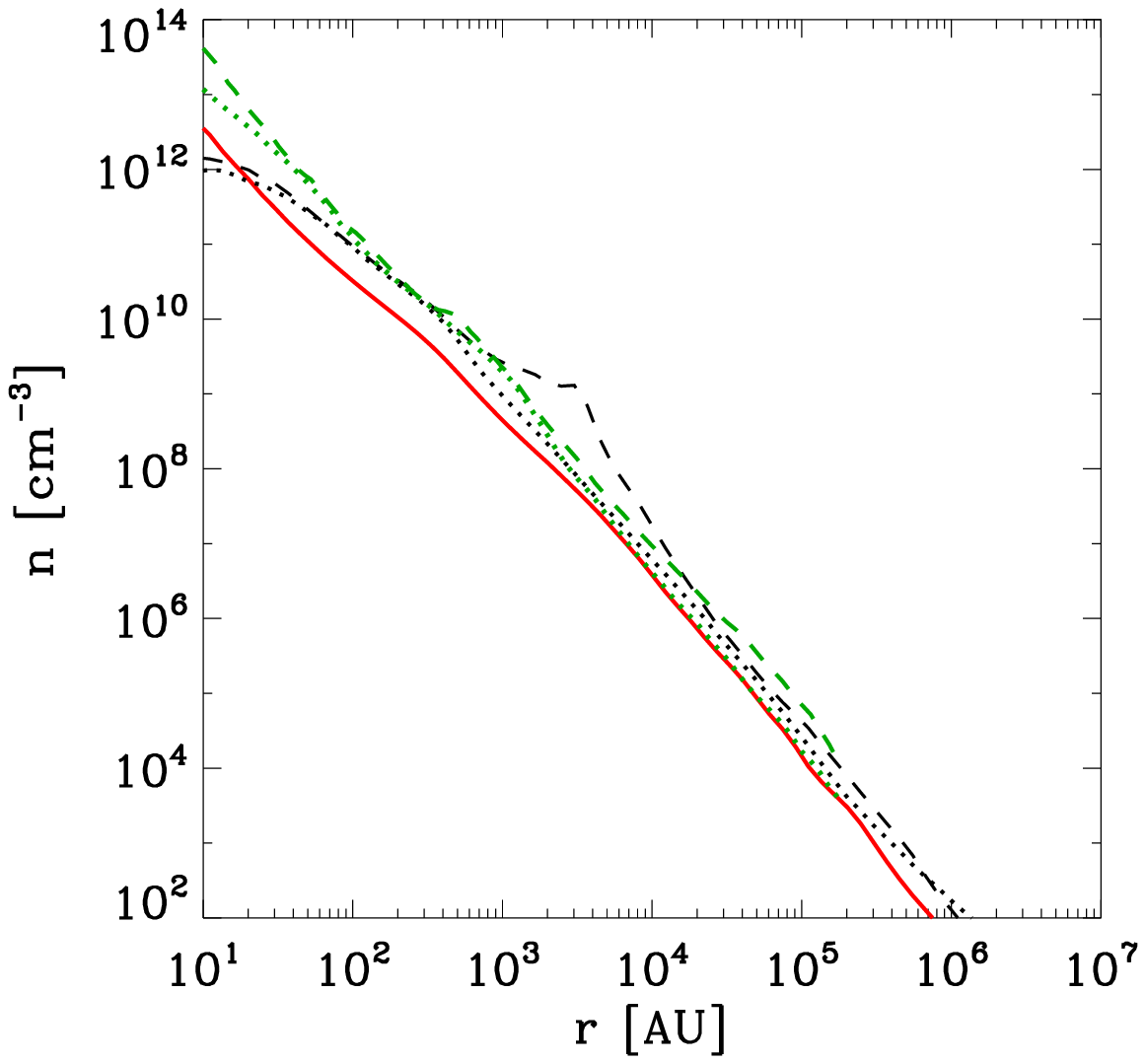}
\includegraphics[width=.4\textwidth]{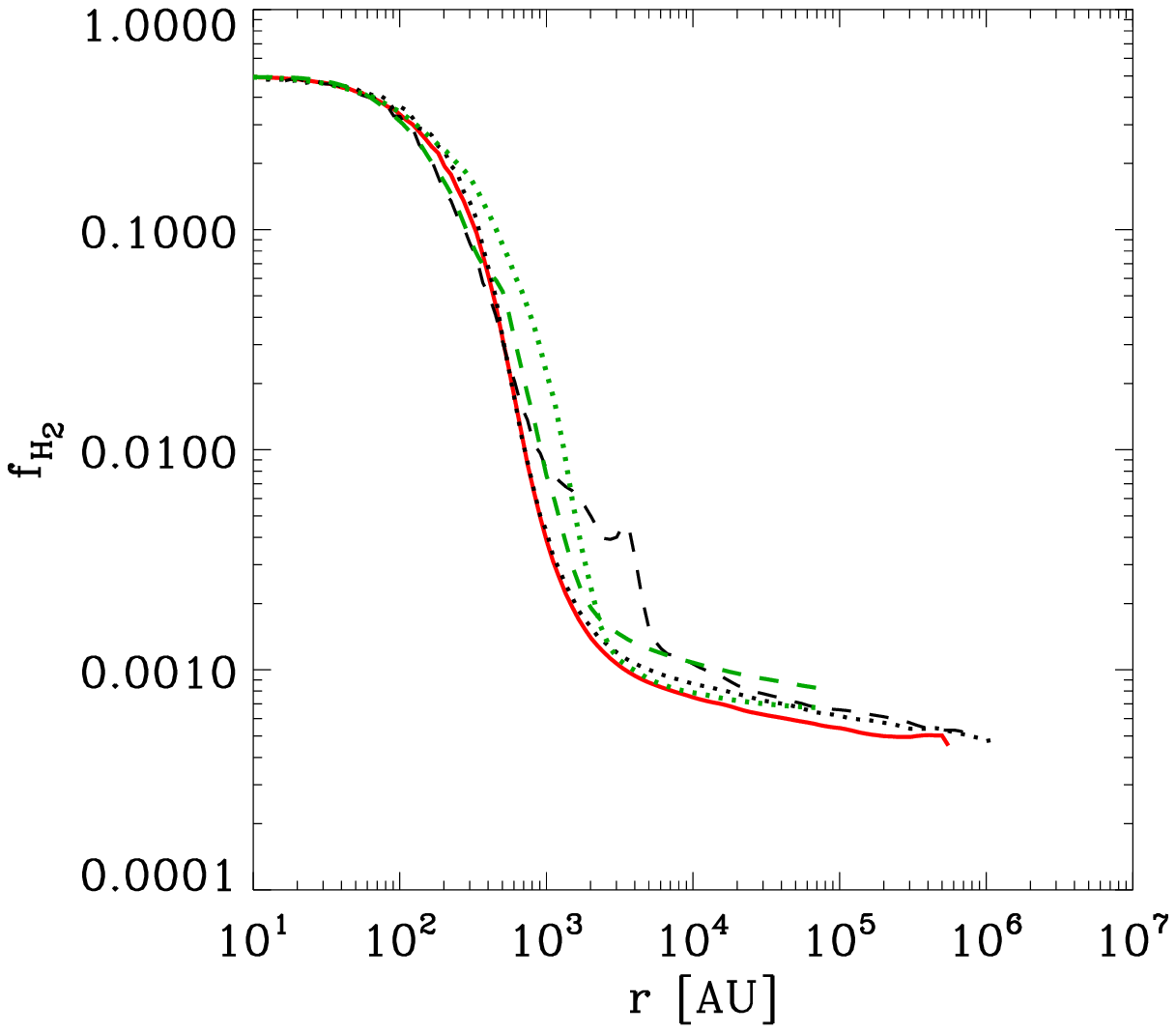}
\end{figure*}
\begin{figure*}
\includegraphics[width=.4\textwidth]{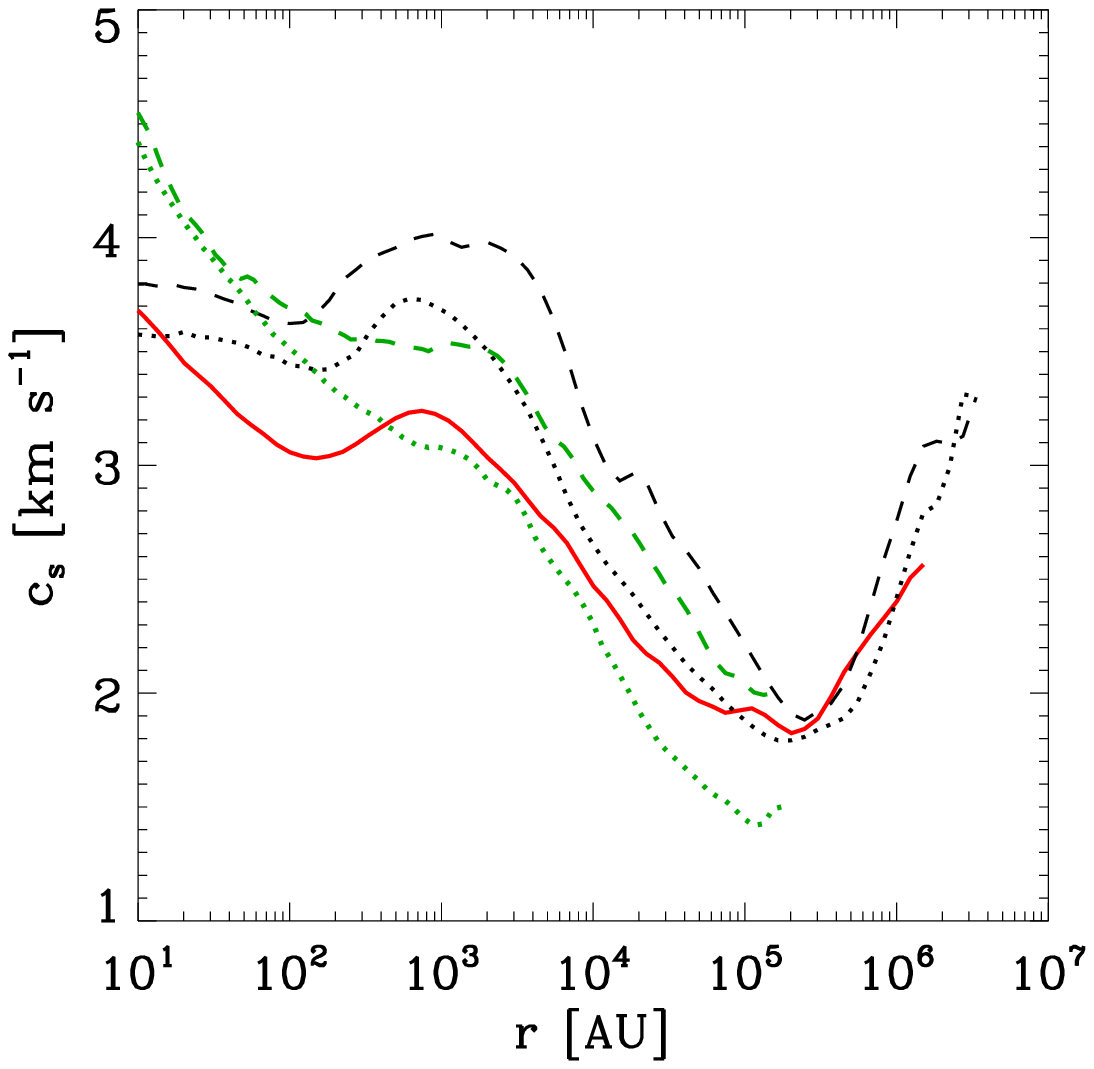}
\includegraphics[width=.4\textwidth]{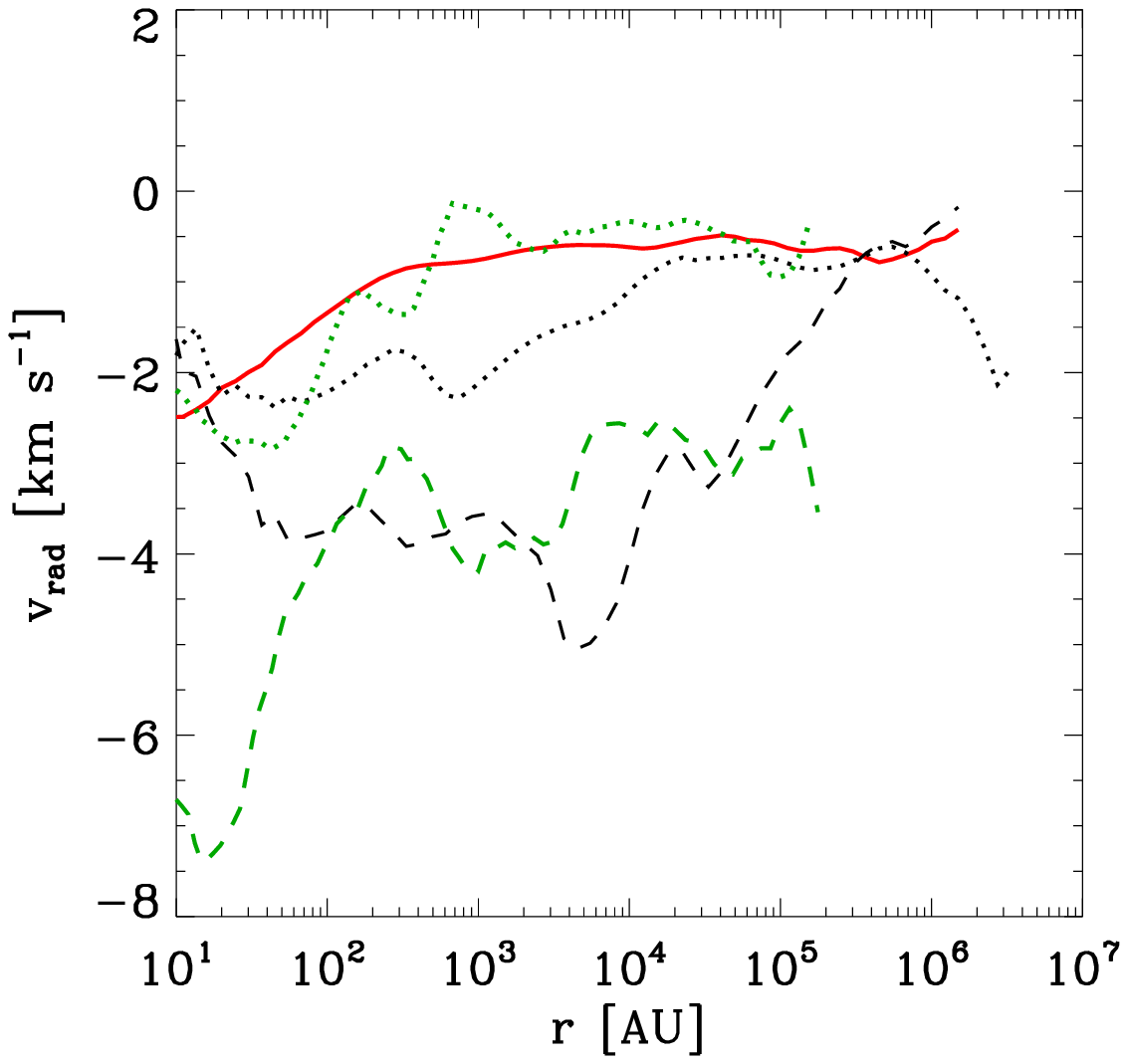}
\caption{
Radial profiles of various gas properties with respect to the highest-density particle, just prior to the initial protostar or sink particle formation.  
In all panels, dashed and dotted lines represent minihalos from simulations presented in Stacy \& Bromm (2013) and Greif et al. (2012), as described in the previous figure.
%Line styles have same meaning as in previous figures.
{\it Upper left:} Number density vs. radius.
{\it Upper Right:} H$_2$ fraction vs. radius.
{\it Lower Left:}  Profile of sound speed $c_{\rm s}$. 
{\it Lower Right:}  Profile of $v_{\rm rad}$.
The combination of smaller sound speed, as well as lower  $|v_{\rm rad}|$ and number density, lead to unusually low accretion rates within our $z=15$ minihalo  as compared with those from other studies.
}
\label{csprof}
\end{figure*}

\subsection{Comparison with Other Minihalos}

\subsubsection{Global Minihalo Characteristics}
As will be discussed in Section 4.3, our simulated Pop III system has an unusually low accretion rate.  
We here examine whether this is due to the characteristics of its host minihalo.  
We first measure the evolution of the virial mass of the minihalo considered here, and compare to four other minihalos taken from the cosmological simulations presented in \cite{greifetal2012} 
and \cite{stacy&bromm2013}, where we have chosen the minihalos hosting the most rapidly and most slowly accreting stellar systems from each of those two studies.  
We determine which particles reside in the halo by first locating the simulation region's densest gas particle, or hydrodynamic mesh element in the case of the {\sc arepo} simulations.
Making the simple assumption that this point marks the center of the
halo, the extent of the halo was determined by finding the surrounding
spherical region in which the average DM density is 200$\rho_b$,
where  $\rho_b \simeq 2.5\times 10^{-30} \left( 1+z \right)^3$ g cm$^{-3}$  
is the redshift-dependent background density.

The minihalo of our simulation has the minimum necessary mass of $M_{\rm halo} \la 10^{6}$ M$_{\odot}$ before gas condensation and H$_2$ cooling begins, and this is very close to the mass of
other Pop III star-forming halos (Fig. \ref{halomass}).
However, it does not reach this minimum $M_{\rm halo}$ until the relatively late time of $z=15$.
%which leads to a slower growth rate of both the DM and baryonic components of the minihalo (Fig. \ref{halomass}).  
By this redshift, Hubble expansion has reduced the density of the background universe, leading to a slower accretion rate from the surrounding cosmic web.
In particular, the average DM accretion rate $\dot{M}_{\rm DM}$ of our minihalo is $3 \times 10^{-3}$ M$_{\odot}$ yr$^{-1}$, as compared to the more typical rate of $10^{-2}$ M$_{\odot}$ yr$^{-1}$ for minihalos which collapsed at $z\sim 30$. Employing this same redshift range and $M_{\rm DM} = 5 \times 10^5$ M$_{\odot}$, we find that our measured  $\dot{M}_{\rm DM}$ values are in excellent agreement with analytical fits determined from other numerical simulations, e.g.

\begin{equation}
\dot{M}_{\rm DM} \simeq 35 \left( 1+z \right)^{2.2} M_{12}^{1.07} {\rm M_{\odot}} {\rm yr^{-1}} \mbox{,}
\end{equation}

\noindent where $ M_{12} = M_{\rm DM}/10^{12}{\rm M_{\odot}}$ (\citealt{geneletal2008}).
%This above formulation 
Though the above estimate was orignally derived from simulations of much larger-mass halos, it also applies accurately to our minihalos. \cite{geneletal2008} furthermore find that the above expression agrees well with the general predictions of extended Press-Schechter (EPS) theory (\citealt{press&schechter1974,mo&white1996,lacey&cole1993}, see also \citealt{neisteinetal2006}). As numerically confirmed in Gao et al. (2005; e.g., their figure 1), the Press-Schechter formalism indeed accurately predicts the accretion rate of minihalos with $M_{\rm halo} \sim 10^6$ M$_{\odot}$. 
They found an accretion rate of $\sim 2.5 \times 10^{-2}$ M$_{\odot}$ yr$^{-1}$ for their $z=50$ minihalo when it was collecting the bulk of its mass, which in turn corresponds well with the above equation's prediction of $\sim 2 \times 10^{-2}$ M$_{\odot}$ yr$^{-1}$.
This further confirms the expected larger growth rates of these rare-peak and high-redshift minihalos (e.g., \citealt{reedetal2005}).

\begin{figure}
\includegraphics[width=.4\textwidth]{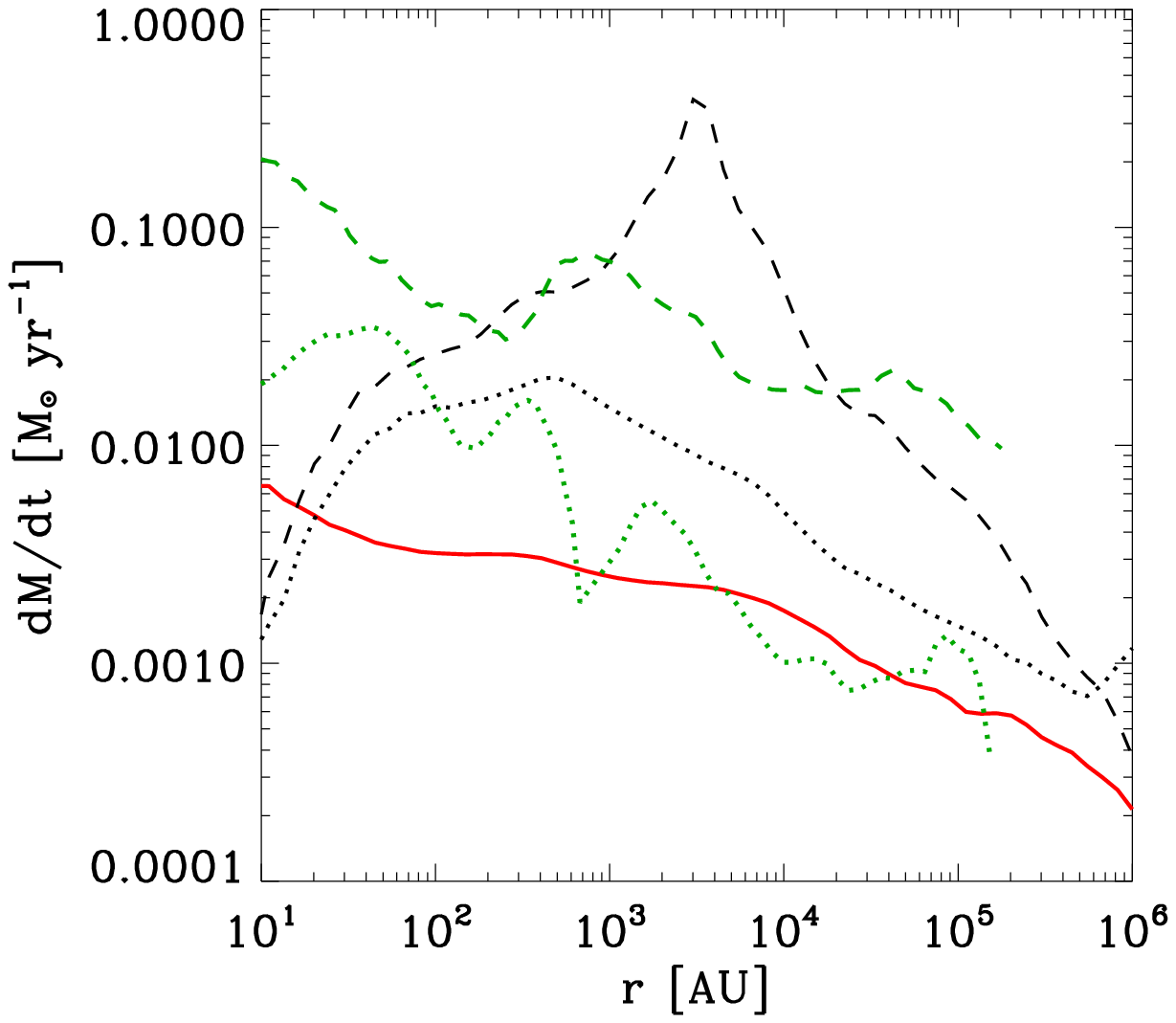}
\includegraphics[width=.4\textwidth]{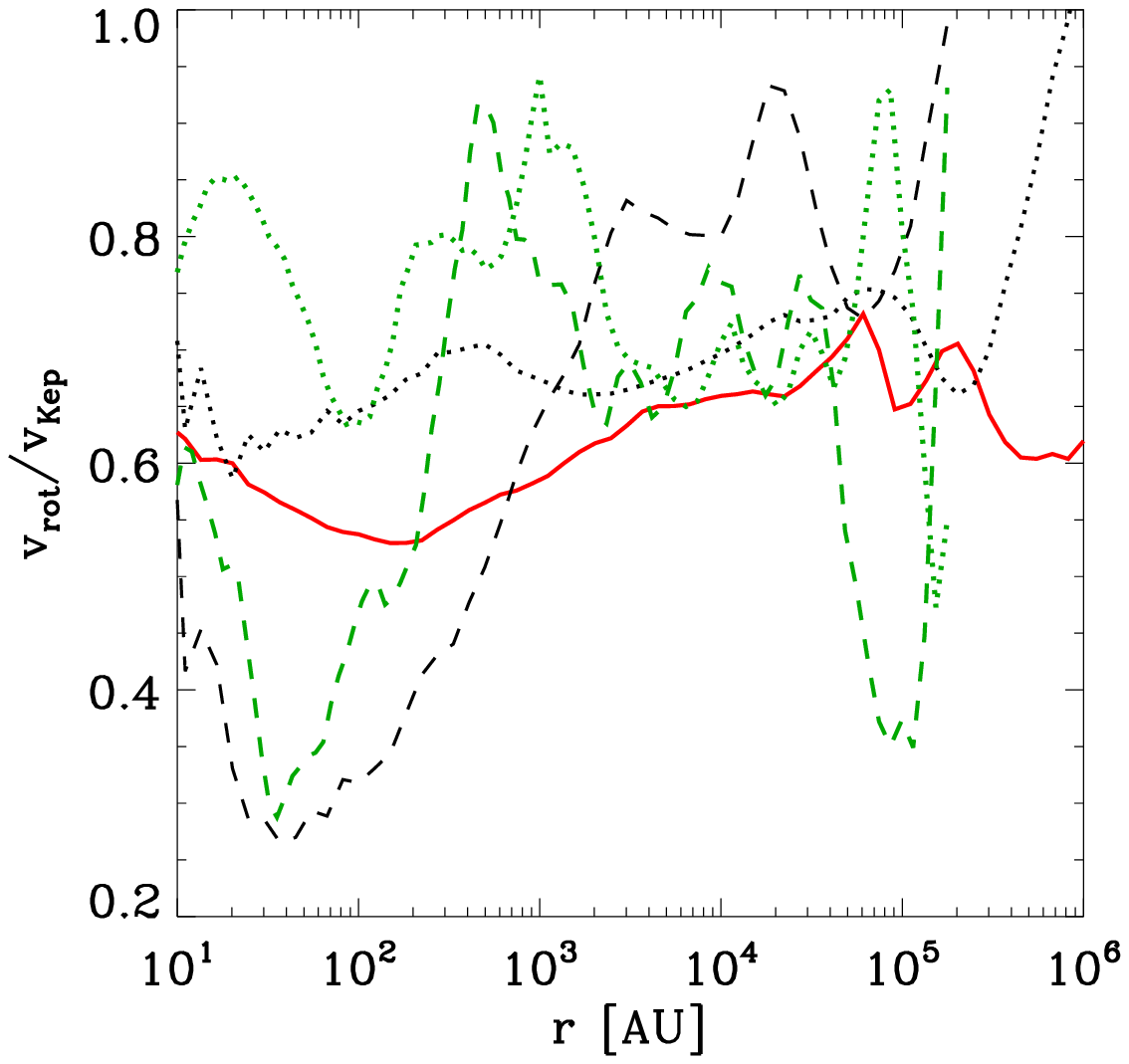}
\caption{
%{\it Left:}
{\it Top:}
Estimated spherical accretion rate $\dot{M}_{\rm sph}$ over a range of distances from densest gas particle, just prior to protostar or sink particle formation.  
Line styles have the same meaning as in previous figures.  Note that the combination of reduced density and $|v_{\rm rad}|$ within the inner few 10$^5$ AU ($\sim$ 0.5 pc) leads to lower accretion rates than typically found in primordial gas.
%{\it Right:}  
{\it Bottom:}
Level of rotational support $f_{\rm rot} = v_{\rm rot}/v_{\rm Kep}$ throughout the gas at this same time.
Note that the two most rapidly accreting systems (green and black dashed lines) are also the only two systems with extended regions of very low rotational support ($f_{\rm rot}$ < 0.5).
}
\label{mdotprof}
\end{figure}

The gas mass within the minihalo, which we take simply as all gas within the halo virial radius with $n>1$ cm$^{-3}$, grows at average rates which range from $M_{\rm gas} = 10^{-4}$ M$_{\odot}$ yr$^{-1}$ for our minihalo to $10^{-3}$ M$_{\odot}$ yr$^{-1}$ for the earlier-collapsing minihalos.  
This may be compared to the growth rate found in previous studies, e.g. 

\begin{equation}
\dot{M}_{\rm gas} \simeq 6.6 \left( 1+z \right)^{2.25} M_{12}^{1.15} f_{0.165}  {\rm M_{\odot}} {\rm yr^{-1}}  \mbox{,}
\
\end{equation}

\noindent where $f_{0.165}$ is the baryonic fraction in the halos in units of the cosmological value $f_{\rm B} =  \Omega_{\rm B} / \Omega_{\rm M} = 0.165$ (e.g., \citealt{dekeletal2009}, see also \citealt{faucher-giguereetal2011}).  Assuming $f_{0.165}=1$ we find that, over the redshift range $z=15-30$, $\dot{M}_{\rm gas}$ ranges from 2 to $7 \times 10^{-4}$ M$_{\odot}$ yr$^{-1}$.  Thus, for a given DM halo mass (e.g. $M_{\rm halo}$ = 10$^6$ M$_{\odot}$),  the rate at which both DM and gas will fall into the gravitational potential well will vary by nearly an order of magnitude over this redshift range. This corresponds to the nearly factor of ten reduction in $\rho_b$ as the universe expands over redshifts $z=30$ to $z=15$.

The variation in collapse redshifts seen in these simulations stems from differences in sizes of the cosmological box that was used.  The \cite{greifetal2012} and \cite{stacy&bromm2013} simulations employed box sizes of 500 kpc and 1.4 Mpc (comoving), and were thus able to capture larger `$\nu \sigma$ peak' fluctuations, where $\sigma$ is the standard deviation in the Gaussian random  field of primordial density fluctuations.   $\Lambda$CDM theory predicts that at $z\sim30$, a 10$^6$ M$_{\odot}$ halo corresponds approximately to a 3$\sigma$ peak (e.g., \citealt{loeb2010, bromm2013}).  Our smaller box size of 200 kpc (comoving) captures only 1-2$\sigma$ peaks, which instead corresponds to a later collapse redshift of $z\sim15$ for a  10$^6$ M$_{\odot}$ halo.  
%Each of these simulations then focused only on the first of their halos to to reach the mass threshold for hosting Pop III star formation. 
%Thus, while 10$^6$ M$_{\odot}$ halos are relatively rare at $z=30$, they are more common by $z=15$.  

\subsubsection{Star-forming Core Characteristics}
Along with slower overall halo gas accretion 
%rates at lower redshift lead to a 
in the $z=15$ minihalo, we find a
reduced gas accretion rate within the central parts of the minihalo as well, which may be roughly estimated through the gas soundspeed $c_{\rm s}$ (Figure \ref{csprof}).
Although we find good agreement in  $c_{\rm s}$  beyond 10$^5$ AU ($\sim$ 0.5 pc), there is divergence in $c_{\rm s}$ in the inner regions.
In our simulation the sound speed of the gas within 10$^4$ AU is $\sim$ 3  km s$^{-1}$.  This is slightly less than that seen in some of the comparison halos, $\sim$ 4 km s$^{-1}$ (Figure \ref{csprof}).  Making the simple estimate that the Jeans mass is infalling at the free-fall rate, we can scale the accretion rate with soundspeed as
$\dot{M} \simeq c_{\rm s}^3/G$, 
similar to the \cite{shu1977} similarity solution for collapse of a singular isothermal sphere.
From this we may predict accretion rates ranging from $\sim 6 \times 10^{-3}$ M$_{\odot}$ yr$^{-1}$ for the $z=15$ minihalo to $\sim 1.5 \times 10^{-2}$ M$_{\odot}$ yr$^{-1}$ for the higher-redshift halos. 

Other properties of the gas, such as the density profile and H$_2$ fraction (lower panels of Figure \ref{csprof}), show interesting variation between minihalos, as well.  
At distances greater than 100 AU, the H$_2$ fraction can vary by approximately an order of magnitude, though in each star-forming cloud the gas is fully molecular within the central 100 AU.  We do not find an exact correlation between H$_2$ fraction and $c_{\rm s}$.
However, our $z=15$ minihalo generally has the lowest H$_2$ fraction at distances greater than 1000 AU as well as nearly the lowest temperatues at all distances.  
The gas density is unusually small as well.
Though for each star-forming core the density roughly follows a $\rho \propto r^{-2}$ profile, beyond 20 AU the $z=15$ minihalo profile is normalized to smaller values than the others.  
At some radii the gas density is an order of magntiude lower than the minihalo with the highest density. 

As shown in Figures \ref{csprof} and \ref{mdotprof}, we may also approximate the spherical accretion rate $\dot{M}_{\rm sph}$ that results from the density and radial velocity profiles, which is appropriate when considering gas within the minihalo where
the density profile is more spherically symmetric.
We estimate $\dot{M}_{\rm sph}$ within a range of radial bins as:

\begin{equation}
\dot{M}_{\rm sph} = 4 \pi  r^2 \rho v_{\rm rad} \mbox{,}
\end{equation}

\noindent where $r$ is the distance as measured from the densest gas particle, and $v_{\rm rad}$ is the average radial velocity of gas within the radial bin.  As is apparent in Figure \ref{csprof}, between 100 AU and 10$^5$ AU (0.5 pc) the gas within our minihalo typically infalls at $v_{\rm rad} \sim $ 1 km s$^{-1}$, several times smaller than  $v_{\rm rad} \sim $ 4 km s$^{-1}$ as seen in the most rapidly growing halos.  
Together with the gas density this yields a gas accretion rate between 100 and 10$^5$ AU that ranges from $\sim 6 \times 10^{-4}$ M$_{\odot}$ yr$^{-1}$ to  $\sim 3 \times 10^{-3}$ M$_{\odot}$ yr$^{-1}$, approximately an order of magnitude less than the accretion rates found in the high-redshift halos.  
For all minihalos,  $\dot{M}_{\rm sph}$ is also substantially lower than the accretion rate estimated from $c_{\rm s}$, indicating that angular momentum support slows gas infall.
We furthermore note that the halo which formed at the second-most recent redshift (green dotted lines in Figures \ref{halomass} to \ref{mdotprof}) has DM and gas accretion rates and $\dot{M}_{\rm sph}$ values which are intermediate between the $z=15$ minihalo and the highest-redshift halos.  

These results may indicate a correlation between collapse redshift and rate of mass infall even on small scales.
However, \cite{gaoetal2007} found that while gas within higher-redshift halos will indeed reach protostellar densities in shorter timescales, there was no subsequent correlation between formation redshift and instantaneous accretion rate at the densities they resolved (10$^{10}$ cm$^{-3}$).
\cite{oshea&norman2007} even find that Pop III star-forming regions which form later have higher maximum accretion rates, the opposite trend to ours.  They attribute this to the increasing virial temperature of halos with redshift, $T_{\rm vir} \propto M^{2/3}_{\rm vir}\left( 1 + z \right)$. In their chain of reasoning, this would lead to warmer overall gas temperatures as the minihalo first collapses, yielding more rapid H$_2$ formation rates.  
The higher H$_2$ fraction in turn would lead to cooler gas in the cores of the halos, and thereby slower accretion rates.  
In Figure \ref{csprof}, however, we find a counter example to this.  A relativly high-redshift minihalo with an enhanced H$_2$ fraction (black dashed line) in fact has the warmest temperatures in both the core and outer region, and in some regions beyond 100 AU, it also has the highest accretion rates.

\cite{gaoetal2007}, on the other hand, emphasize the importance of angular momentum.  They find that as gas collapses its properties become independent of the global properties of the halo, and that more disk-like and rotationally supported inner star-forming clouds will have lower accretion rates.  
In the bottom panel of Figure \ref{mdotprof} we compare the levels of rotational support throughout the halos, defined as $f_{\rm rot}$ = $v_{\rm rot}$/$v_{\rm Kep}$, where $v_{\rm Kep} = \left(G M_{\rm enc}/r \right)^{1/2}$. 
The $z=15$ minihalo has  $f_{\rm rot} \sim 0.6$ over several orders of magnitude in distance.  The other two more slowly-accreting systems (green and black dotted lines in Figure \ref{mdotprof}) also maintain high  $f_{\rm rot}$ between 10 and 10$^6$ AU, with $f_{\rm rot}$ consistently greater than 0.6.  In contrast, the two most rapidly accreting systems (green and black dashed lines) have extended regions where  $f_{\rm rot}$ falls below 0.5.  Thus, similar to the conclusions of \cite{gaoetal2007}, angular momentum structure, combined with that of temperature and density, plays a key role in the rate of gas infall.  

\begin{figure}
\includegraphics[width=.45\textwidth]{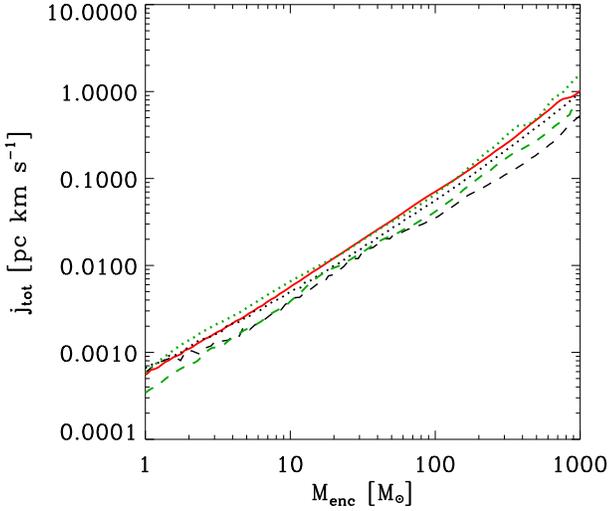}
\caption{
Angular momentum profiles from various simulated minihalos, taken at the point of first sink or protostar formation.
Solid red line is taken from the simulation discussed here.
Dashed black line is the most rapidly accreting cluster from Stacy \& Bromm (2013).
Dotted black line is the most slowly accreting cluster from Stacy \& Bromm (2013).
Green dotted and dashed
lines are taken from Greif et al.
(2012) minihalos which had the lowest and highest accretion rates, respectively.
%Solid red line represents the work of Stacy et al. (2010),
%the dotted red line Yoshida et al. (2006), and
%the dashed red line Abel et al. (2002).
}
\label{amomprof}
\end{figure}

Despite variation in level of rotational support, the specific angular momentum profile of the central gas for each minihalo we examine, shown in Figure \ref{amomprof}, all follow a similar $j_{\rm tot} \propto M_{\rm enc}$ powerlaw.  However,  it is interesting to note that the profiles of two most rapidly accreting halos (green and black dashed lines in Figure \ref{amomprof}) are normalized at nearly a factor of two lower than the minihalo of our simulation.   The greater rates at which the central gas flows inward (Figure \ref{csprof}), 
as well as the reduced rotational support (Figure \ref{mdotprof}, bottom panel),
indicates that in these halos $v_{\rm rad}$ dominates over ${v}_{\rm rot}$, 
leading to higher accretion rates at the point of protostellar formation. 

\begin{figure*}
\includegraphics[width=.4\textwidth]{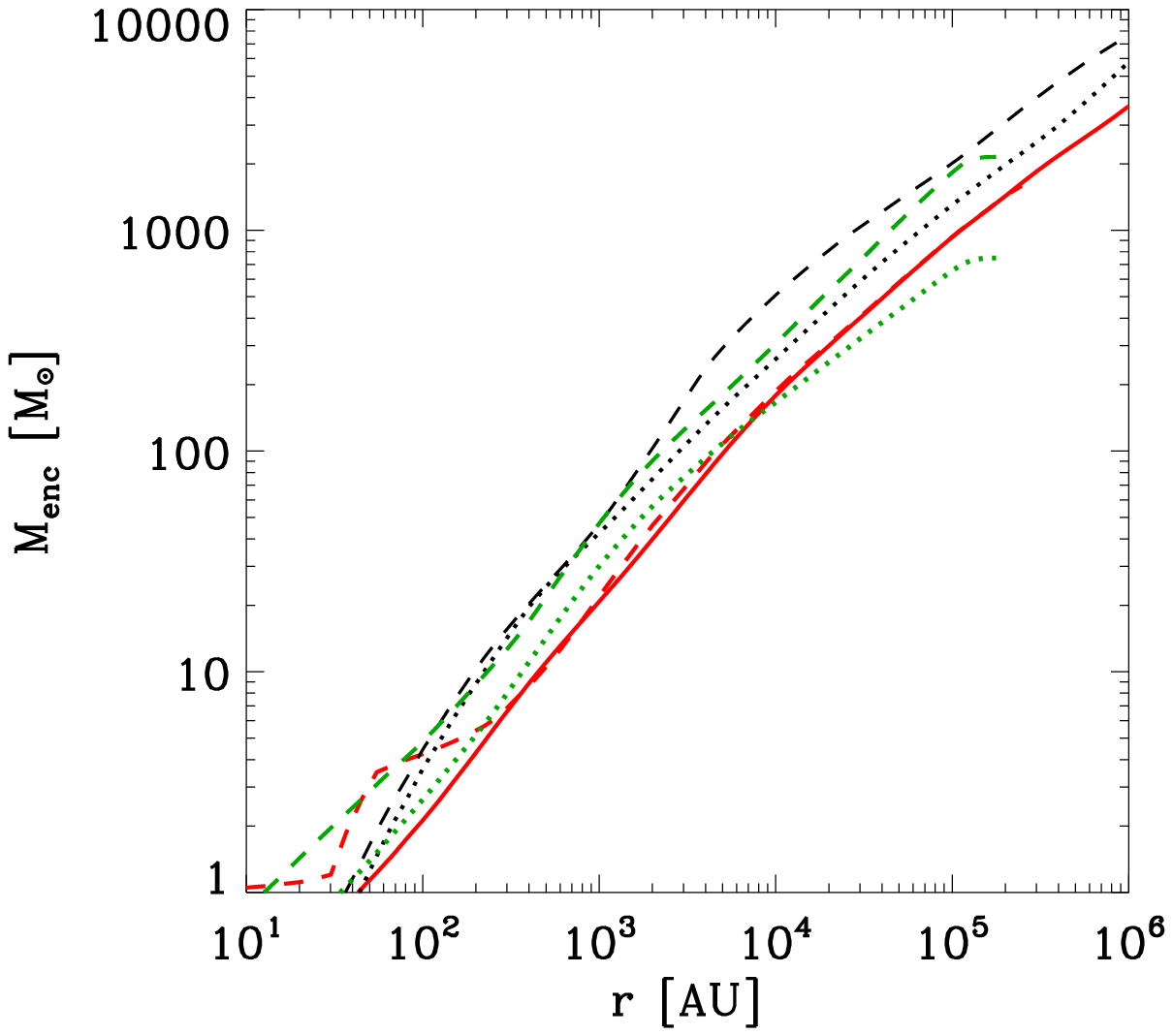}
\includegraphics[width=.4\textwidth]{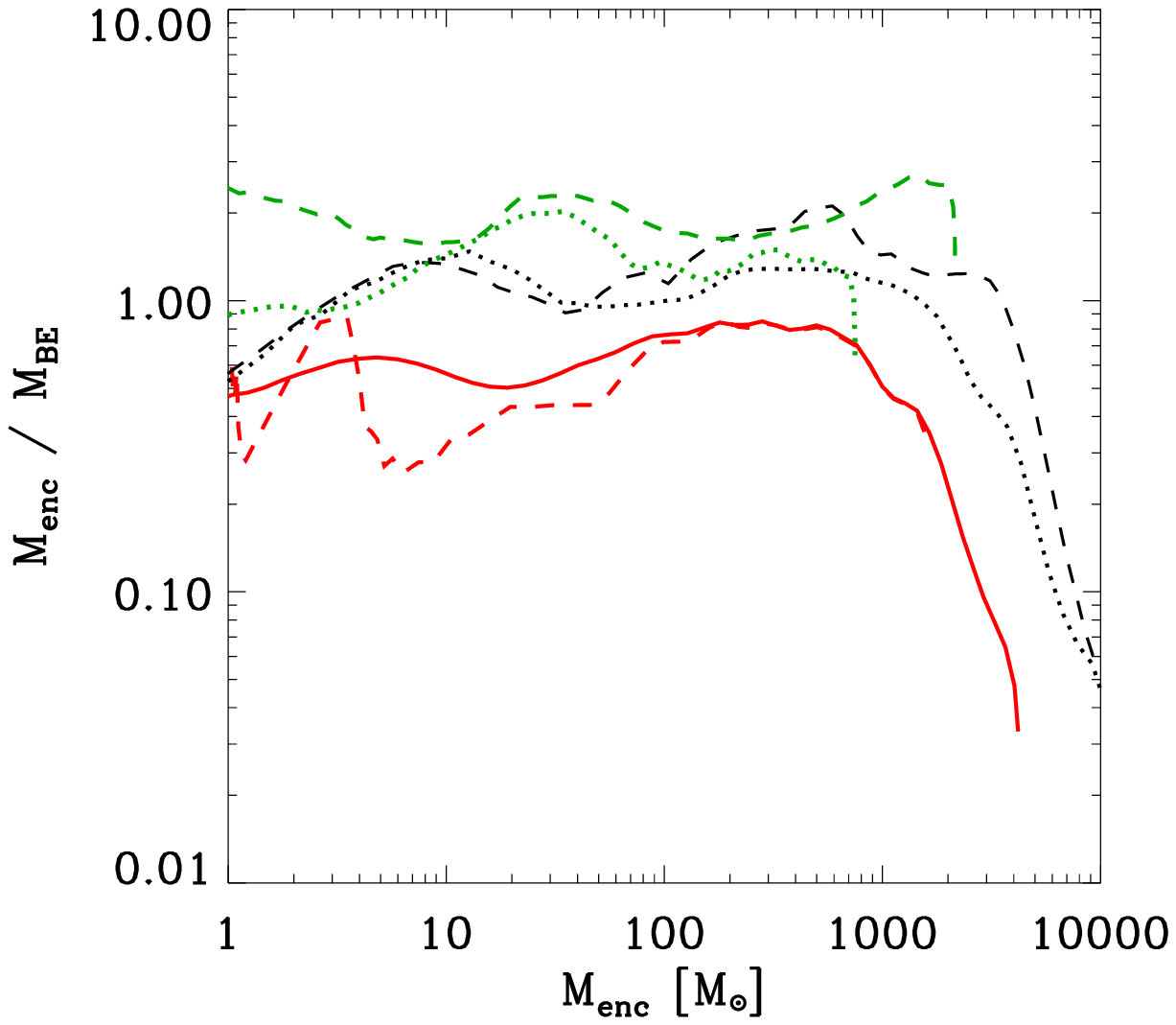}
\caption{
{\it Left:} Enclosed mass $M_{\rm enc}$ versus radius for various minihalos at the point of first sink or protostar formation.
{\it Right:} Ratio of  $M_{\rm enc}$  to Bonnor-Ebert mass $M_{\rm BE}$ versus  $M_{\rm enc}$. 
Solid red line represents the simulation presented in this work just prior to initial sink formation, while the dashed red line represents this simulation 5000 yr after sink formation.
Other lines have the same meanings as in previous figures.  
Central point is taken as the most dense gas particle or most massive sink.
Minihalos with higher overall accretion rates and less rotational support have lower overall $M_{\rm enc}$ at a given radius and a lower ratio of $M_{\rm enc}$ to  $M_{\rm BE}$.
}
\label{menc}
\end{figure*}

As expected, the mass enclosed within a given radius is also larger for those minihalos with less rotational support.  The resulting ratio of enclosed mass $M_{\rm enc}$ to Bonnor-Ebert mass $M_{\rm BE}$ is larger as well (Fig. \ref{menc}).  
We estimate $M_{\rm BE}$ as

\begin{equation}
M_{\rm BE} \simeq 1000 {\rm M_{\odot}} \left(\frac{T}{200{\rm K}}\right)^{3/2} \left(\frac{n}{10^{4}{\rm cm^{-3}}}\right)^{-1/2} \mbox{.}
\end{equation}

\noindent $M_{\rm BE}$ is similar to the Jeans mass, and a ratio of $M_{\rm enc}$ to $M_{\rm BE}$ that is close to unity roughly indicates gravitational instability of the gas.  

We note that \cite{desouzaetal2013} also emphasize the influence of rotation on the Pop III IMF, finding from their cosmological simulation that the spin distribution of gas within minihalos evolves with redshift.  They employed the model from \cite{mckee&tan2008} which used semi-analytic methods to find a relation between gas rotational support, effectiveness of protostellar feedback, and ultimate protostellar mass.  Assuming one star per minihalo, \cite{desouzaetal2013} determine that the Pop III IMF should evolve to have lower peak masses at lower redshift.  However, they point out that correlating spin with a Pop III IMF will be complicated by further feedback effects and Pop III multiplicity.

\begin{figure}
\includegraphics[width=.45\textwidth]{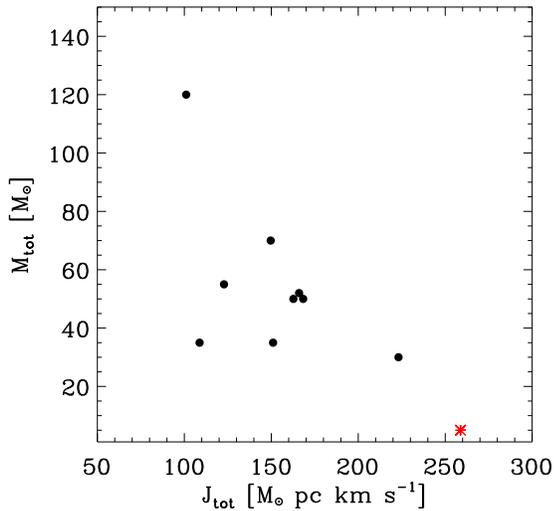}
\caption{Total mass accreted onto the stellar system after 5000 yr versus the total angular momentum within the central 200 M$_{\odot}$ of enclosed mass.  Angular momentum is measured just before the first sink forms.  Values are taken from this simulation (red asterisk) as well as the minihalos studied in Stacy \& Bromm (2013, filled circles).  Note that the gas with the highest angular momentum also has the lowest stellar mass and accretion rate.
}
\label{amom-vs-mass}
\end{figure}

Our results concerning the relation between angular momentum and protostellar accretion rate agree with the those of \cite{gaoetal2007}.  The implication that the IMF may shift to lower mass with lower redshift also shows a rough agreement with \cite{desouzaetal2013}, but for different physical reasons.
However, neither these authors nor \cite{oshea&norman2007}  simulated the evolving accretion rate of the protostellar cloud after the first protostar formed.  Our study allows for this through the sink particle method.    
As will be further discussed in Section 4.3, we find that the slower and more rotationally-dominated gas infall within the $z=15$ minihalo later leads to reduced accretion rates on to the evolving stellar system as well.  In Figure \ref{amom-vs-mass} we compare total sink mass after 5000 yr and the total angular momentum within the central 200 M$_{\odot}$ just prior to sink formation. We use the minihalo of this work as well as the suite of ten minihalos presented in \cite{stacy&bromm2013}.  Some anti-correlation is apparent, particularly in that the $z=15$ minihalo has both the lowest total stellar mass and the highest central angular momentum.  
However, this trend does have significant scatter since other processes such as turbulent angular momentum transport and N-body dynamics will affect the growth rate of the stellar cluster.

\begin{figure*}
\includegraphics[width=.8\textwidth]{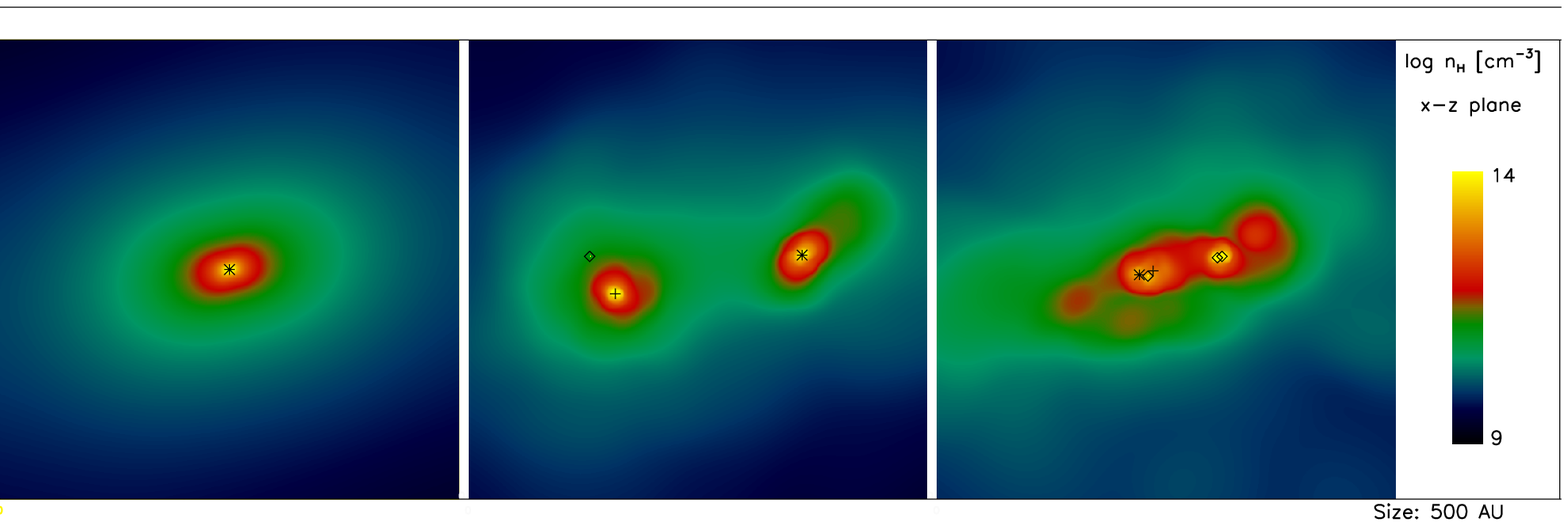}
\includegraphics[width=.8\textwidth]{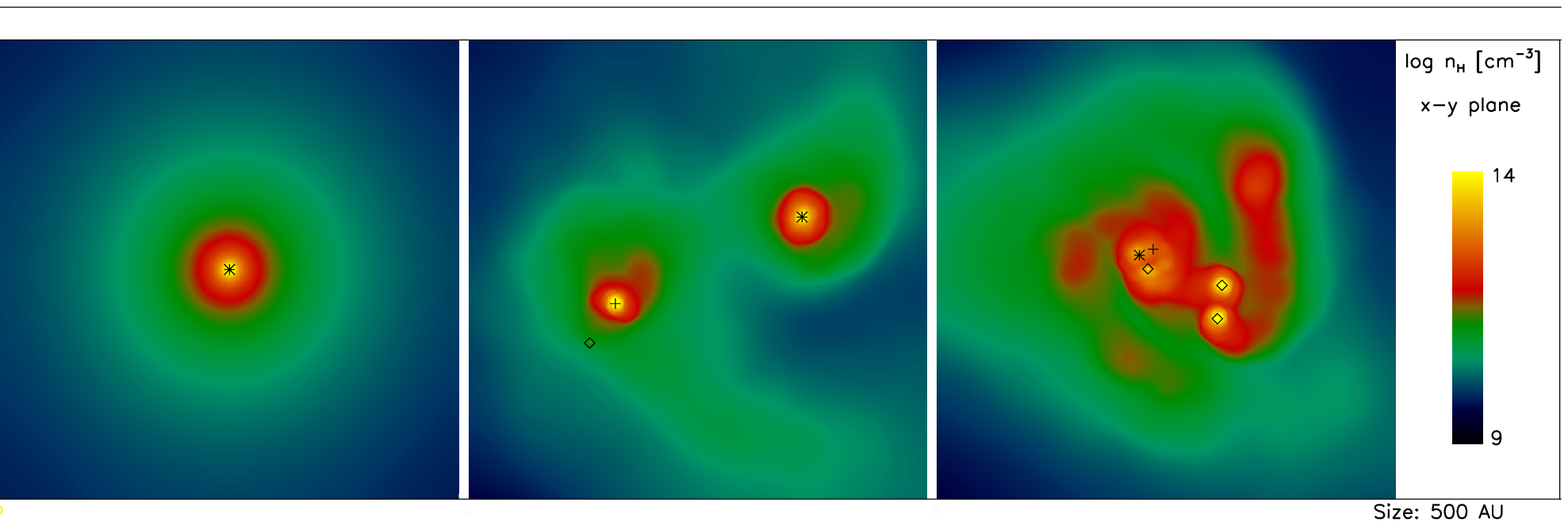}
 \caption{Density projection of central 500 AU of gas around first sink.  
%Top row is the `no-feedback' case in the x-z plane, 
Top and bottom rows are the x-z and x-y planes, respectively.  From left to right, times after sink formation are 3, 2000, and 3500 yr.  
Asterisk is the largest sink.  Plus symbol is the second largest sink.  Diamonds are other secondary sinks.
Note the rapid changes in the sink orbital motion and the structure of the protostellar disk.}
\label{nh-morph}
\end{figure*}

\subsection{Evolution of Protostellar Disk}

As the gas collapses and begins to form sink particles, it also develops into a flattened disk structure 
(Figure \ref{nh-morph}).
We may estimate the dependence of the mass inflow rate inside the accretion disk on density and temperature using the following:

\begin{equation}
 \dot{M}_{\rm disk} = 3\pi \nu \Sigma \mbox{,}
\end{equation}

\noindent where $\Sigma$ is the disk surface density and $\nu$ is estimated based upon the prescription introduced by \cite{shakura&sunyaev1973},

\begin{equation}
 \nu = \alpha_{\rm SS} H_{\rm p} c_{\rm s} \mbox{.}
\end{equation}

\begin{figure}
\includegraphics[width=.45\textwidth]{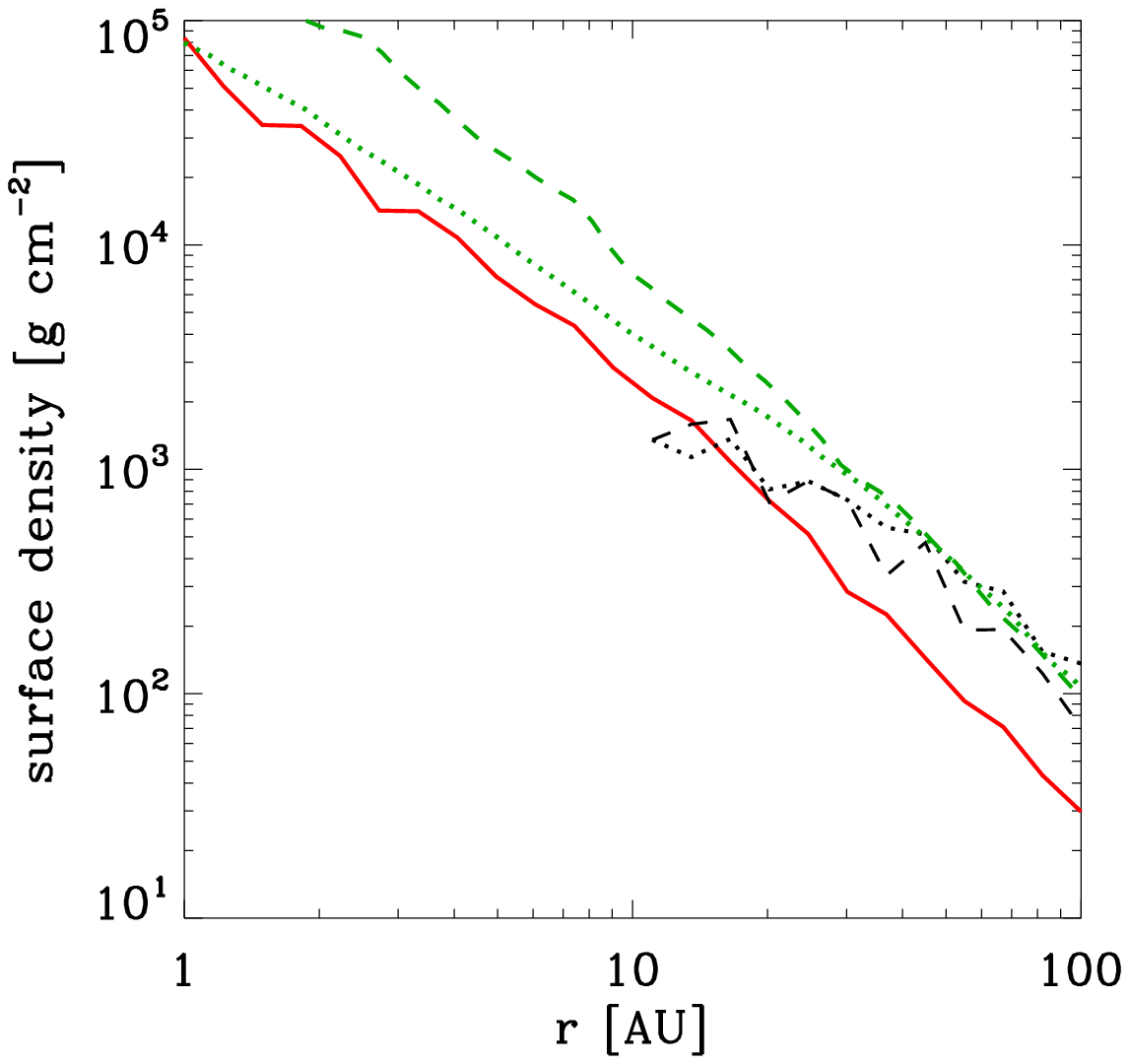}
\caption{
Surface density profile of this simulation (solid red line) as well as two selected minihalos from Stacy \& Bromm (2013; black dotted and dashed lines) and two from Greif et al (2012, green lines).  
Profiles are shown just prior to the initial sink formation.  Note that the Stacy \& Bromm (2013) calculations were resolved down to $\sim$ 10 AU, so we cannot calculate their surface densities on scales smaller than this.  
%The black dotted line represents the Stacy & Bromm (2013) minihalo which had the lowest overall stellar accretion rate, while the black dashed line represents that which had the greatest accretion rate.
%Also shown is the surface density profile from the simulation described in Clark et al. 2011 (dash-dotted line).
}
\label{densprof}
\end{figure}

\noindent $H_{\rm p}$ is the pressure scale height of the disk, and $\alpha_{\rm SS}$ is a non-dimensional parameter ranging between $\sim 10^{-2}$ and 1, depending on the nature of angular momentum transport in the disk.  
In Figure \ref{densprof} we show the inner surface density profile as measured within a thin 20 AU slice through the central disk, measured just prior to the formation of the first sink particle.  We compare with the same set of other minihalos as previously discussed.  Note that the length of 20 AU is chosen because this is the resolution limit of the comparison minihalos first presented in \cite{stacy&bromm2013}.  Even on scales as small as 100 AU, the disk has surface densities as much as three times lower than the comparison minihalos.  From Equation 16 we see that this leads to a correspondingly reduced disk accretion rate which persists well after the first sink appears.  

Let us roughly estimate $\alpha_{\rm SS}\simeq 0.1$,
and $c_s\sim$ 3 km s$^{-1}$, corresponding to a temperature of $\sim$ 1000 K. 
The disk scale height at $R=100$ AU may be estimated as $H_{\rm p}/R \sim c_s/v_{\rm rot}(R)$.  From Fig. \ref{velprof} we estimate $v_{\rm rot}(R=100 \rm AU) \sim 3$ km s$^{-1}$ and thus  $H_{\rm p}\sim 100 AU$.    
We then find 
$\nu \sim 5\times 10^{19}$ cm$^2$ s$^{-1}$.  

The surface density at 100 AU ranges from 30 to 100  g cm$^{-1}$, yielding $ \dot{M}_{\rm disk}$ rates which range from 
$2 \times 10^{-4}$ to $6 \times 10^{-4}$  M$_{\odot}$ yr$^{-1}$.  
Assuming the sinks grow at similar rates, we would expect  after 10,000 yr that the stellar systems would reach a total mass of 
2-6 M$_{\odot}$. 
The lower end of this mass range is within good agreement with the total mass accretion rate seen in the simulation.  However, the upper end is still somewhat lower than the total sink masses found in other calculations (Fig. \ref{amom-vs-mass}). Further variations in $\dot{M}_{\rm disk}$ are likely to come from differences in $c_s$ (i.e. warmer gas temperatures) and $\alpha_{\rm SS}$.  This also indicates the approximative nature of our calculation. 
%and the significant increases in $\dot{M}_{\rm disk}$ over time in some simulations.  

\begin{figure}
\includegraphics[width=.45\textwidth]{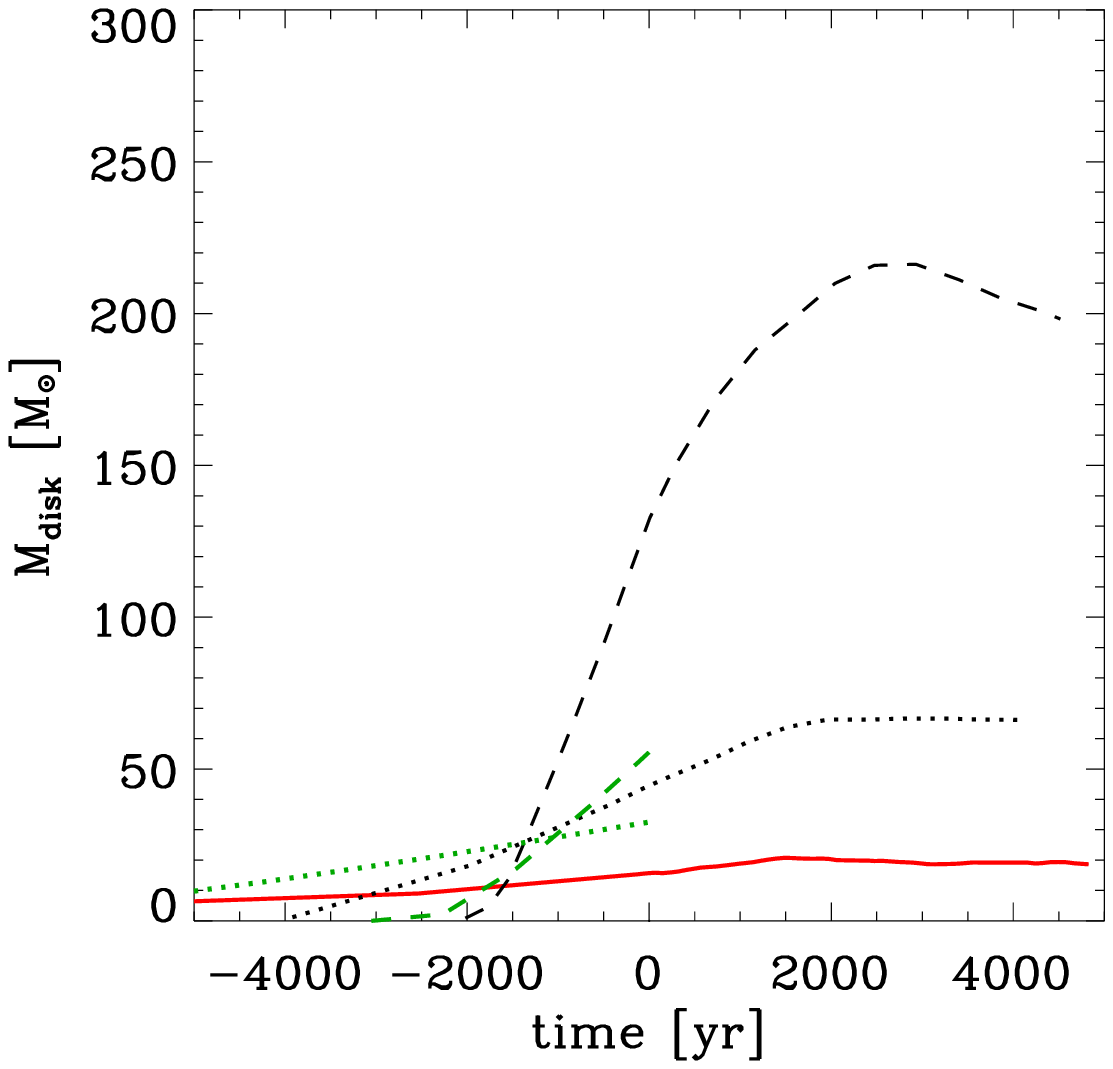}
 \caption{
 Evolution of disk mass over time.  Time is measured with respect to the point at which a sink or protostar first forms.  The solid red line represents the disk mass for our $z=15$ minihalo.
%, while the dashed red line represents the mass of its dense warm phase. 
The black dotted line represents the minihalo from Stacy \& Bromm (2013) that had the lowest overall stellar accretion rate, while the black dashed line represents that which had the greatest accretion rate.
Green dotted and dashed lines additionally show the disk mass within two minihalos from Greif et al. (2012) which had the lowest and highest accretion rates, respectively.  
 }
\label{diskmass}
\end{figure}

The range of accretion rates between the stellar disks also widens over time, such that after 5000 yr the total disk mass of the fastest-accreting halo from \cite{stacy&bromm2013} is a factor of ten greater than the total disk mass within our $z=15$ minihalo.  To illustrate this,  in Figure \ref{diskmass} we show the resulting evolution of the disk mass, as well as the mass of disks taken from the simulations of \cite{stacy&bromm2013} and \cite{greifetal2012}.  In this figure, time is measured with respect to when the first sink forms.   Note that the disk evolution of the green lines does not extend beyond a time of zero because these were taken from simulations which did not form sinks.
The particular point that marks the transition from the stellar disk to the outer envelope is somewhat ambiguous, so to determine whether a gas particle is part of the disk we choose the simple criterion that it must have $n>10^9$ cm$^{-3}$ and $f_{\rm H_2} > 10^{-3}$.  Therefore, only dense and molecular gas is included.  
Compared to other studies, the $z=15$ stellar disk grows at a very low rate.  The slow overall growth rate of the mass of minihalo gas has translated to a slowly growing disk as well.  The second-most slowly growing disk (green-dotted line in Figure \ref{diskmass}) similarly belongs to the second-most slowly growing comparison minihalo.

We briefly note that as mass falls onto the sink, a portion of the gas gets heated to the virial temperature $T_{\rm vir}$ of the sink through release of gravitational potential (Fig. \ref{stuff-vs-nh}).  Given a sink mass of 0.2 M$_{\odot}$ and  using the accretion radius of 1.0 AU, we find a temperature of

\begin{displaymath}
T_{\rm vir}\simeq \frac{G M_{\rm sink}m_{\rm H}}{k_{\rm B}r_{\rm acc}}
\simeq 10^4 \mbox{\,K.}
\end{displaymath}

\noindent  
%Once this gas is heated to several thousand Kelvin, the LW radiation of the sink prevents subsequent re-formation of H$_2$ and cooling of the gas.  
This `warm bubble' of neutral gas expands at its sound speed of $c_{\rm s} \la 10$ km s$^{-1}$.  
Thus, by  $\sim$ 5000 yr the warm bubble has reached a distances of approximately $c_{\rm s} t = 10,000$ AU, which corresponds to gas of density $n \sim 10^7$ cm$^{-3}$.  This warm phase of gas is visible in Figure \ref{stuff-vs-nh} (see \citealt{turketal2010} for further discussion of this warm and neutral gas phase).

These disk properties lead to very low sink accretion rates (Fig. \ref{sinkmass}).  
The disk gas fragments to form a second and third sink 600 and 900 yr after the first sink has formed, and two more between 3000 and 4000 yr.  During the simulation two other sinks form but quickly merge with the initial sink.
A least-squares power-low fit to the growth rate of the three largest sinks remaining at $t_{\rm acc} \sim 5000$ yr yields 

\begin{equation}
M_{*,1} \simeq  0.61 \,{\rm M}_{\odot} \left(t / {\rm 1000 yr} \right)^{ 0.23} 
\end{equation}
\begin{equation}
M_{*,2} \simeq  0.51 \,{\rm M}_{\odot} \left(t /{\rm 1000 yr} \right)^{ 0.28}
\end{equation}
\begin{equation} 
M_{*,3} \simeq 0.62 \,{\rm M}_{\odot} \left(t /{\rm 1000 yr} \right)^{ 0.32} \\
\end{equation}

\noindent Extending these power laws to 1 Myr, a typical accretion time for low-mass stars, these sinks would then reach 
3.1, 3.5, and  5.7 M$_{\odot}$.  A similar approximation for the growth of the least-massive sink predicts a late-time mass of 
0.4 M$_{\odot}$.

 It is uncertain how many more protostars will form at later times and what masses they would reach, since
we do not follow the simulation for sufficient time to track the longer-term evolution of the disk and surrounding $\sim 1000$ M$_{\odot}$ core as the protostars grow. 
%However, it seems unlikely that later gas fragmentation will lead to the collapse of a high-mass star within this core.  
As can be seen in Figure \ref{menc}, the ratio of $M_{\rm enc}$ to $M_{\rm BE}$ has a peak at  $M_{\rm enc} \la 1000$ M$_{\odot}$ and then a rapid drop-off on larger scales.  A similar drop off is seen at greater $M_{\rm enc}$ for the more rapidly accreting halos.  This central several hundred solar masses of material gravitationally infalls toward the central regions at a relatively slow rate of $\la 10^{-3}$ M$_{\odot}$ yr$^{-1}$, leading to little change in the outer $M_{\rm enc}$ profile after 5000 yr (see red dashed lines in Figure \ref{menc}).

At the same time, the protostars followed in our simulation are not projected to become massive enough to develop an HII region that will blow away the gas.  However, the luminosity and LW emission of the protostars will still serve to heat and stabilize the central gas, and the warm phase of the dense gas will continue to grow (Fig. \ref{stuff-vs-nh}).
%, dashed red line in Fig. \ref{diskmass}).  
We define the dense warm phase as gas which has 
$n>10^9$ cm$^{-3}$ and $f_{\rm H_2} < 10^{-3}$, 
%$n>10^6$ cm$^{-3}$ and $T > 4000$ K,
such that only non-molecular gas is included.  This phase approximately consists of $\sim$ 2 M$_{\odot}$ of gas by the end of the simulation.
The diversion of gas to the warm phase instead of the cool disk also helps to explain why $M_{\rm disk}$ does not continue to increase above $\sim$ 20 M$_{\odot}$ after $\sim$ 2000 yr of sink accretion. 
Even with such feedback effects, however, we cannot rule out the possiblity that over the subsequent 10$^5$ to 10$^6$ yr, gas inflow onto the disk will continue until a star reaches $\sim$ 10 M$_{\odot}$ and ionizes its surroundings.  We conjecture that stellar masses greater than 10 M$_{\odot}$ will not be necessary to halt slow inflow like that seen in our simulation, and that even if we followed our simulation for very long times we still would not see a Pop III star reach the more typical masses of 50-100 M$_{\odot}$.  This will be confirmed with future numerical work.

\subsection{Evolution of Stellar orbits}

%Roche Lobe approximation
During the disk evolution, the distance of the secondary sinks from the main sink ranges from $\sim$ 1 AU to a few hundred AU.  
In comparison, the occurrence of Roche-lobe overflow requires the size of one of the binary members to exceed its Roche-lobe radius $r_L$:

\begin{equation}
\frac{r_L}{a} = {\rm max} \left[  0.46224\left(\frac{q}{1+q}\right)^{1/3}, ~ 0.38 \,+ \, 0.2 \,{\rm log}_{10} q \right] 
\end{equation} 

\noindent for $0 < q < 0.8$ (\citealt{paczynski1971}), where $q$ is the binary mass ratio, and $a$ is the semi-major axis.  For $q = 0.1$ we have $r_L = 0.2\,a$, while $q=1$ yields $r_L = 0.38 \,a$.  This brackets the expected range of mass ratios for our simulated binary system.  For $a$ as small as 1 AU, we have $r_L \sim 3-6 \times 10^{12}$ cm, or 40-80 R$_{\odot}$.  As AGB stars can reach well over 100 R$_{\odot}$, this highlights the possibility that an AGB Pop III star may transfer mass to its companion.  

\begin{figure*}
\includegraphics[width=.4\textwidth]{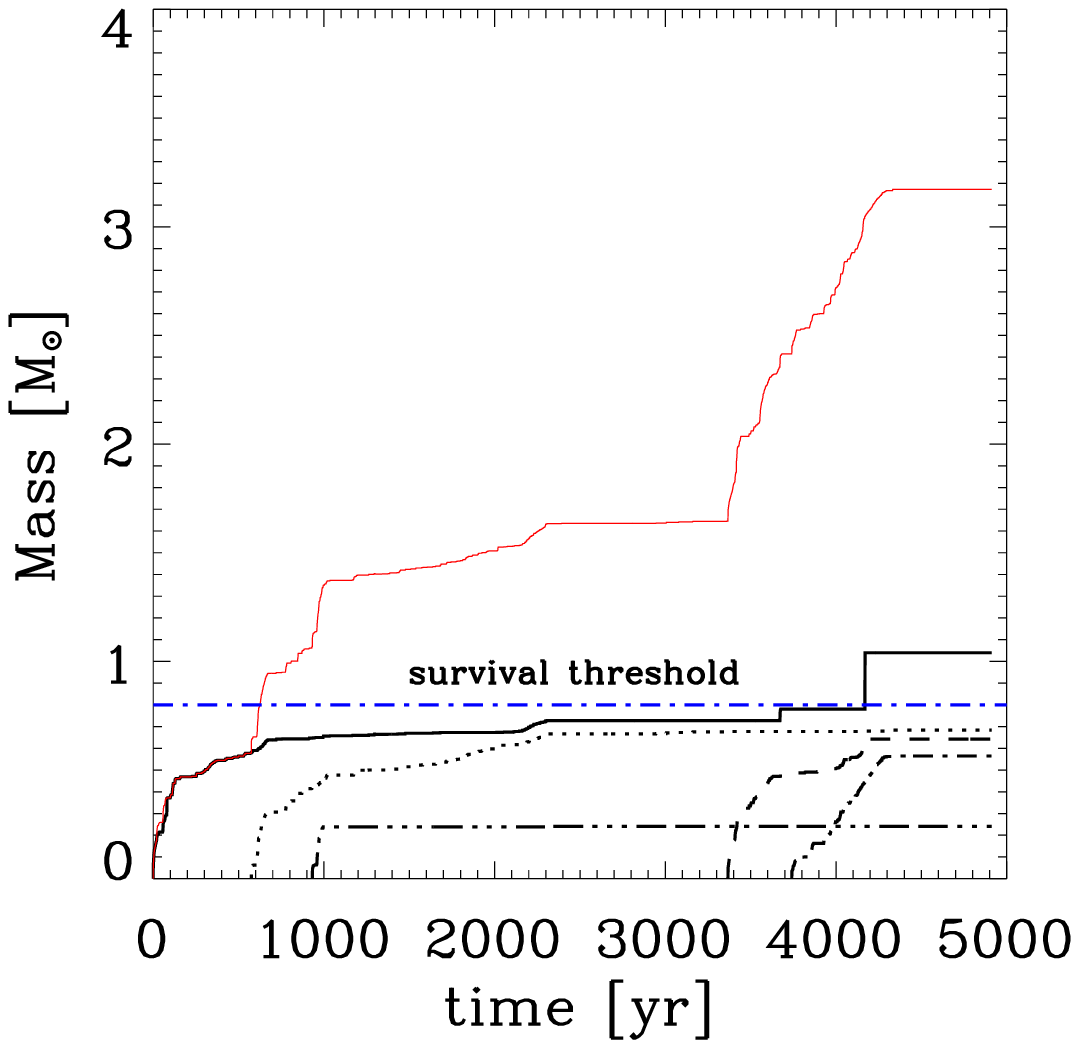}
\includegraphics[width=.4\textwidth]{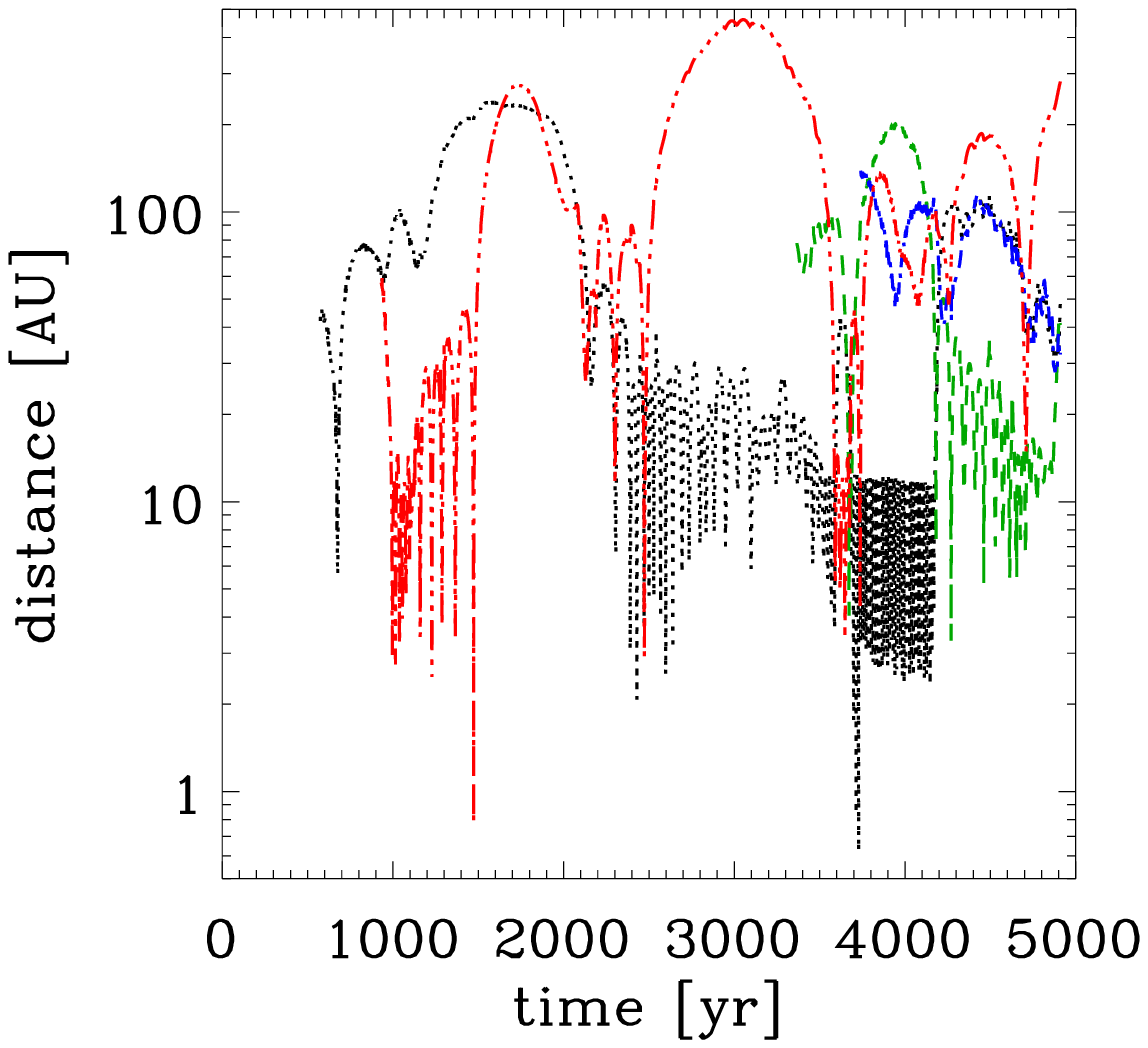}
 \caption{
{\it Left:} Sink growth over time.  Solid line represents the first and largest sink.  Dotted line is the growth of the second-largest sink, while the other three black lines depict the growth of the three remaining sinks that survive to the end of our simulation.  The red line depicts the total sink mass over time.  Blue dash-dot line depicts the `survival threshold', the maximum mass for a star that could survive to the present-day.  
{\it Right:} Distance of secondary sinks from the most massive sink over time.  Line styles refer to the same sinks as in the left panel, but with different colors for more visible contrast between lines.
 }
\label{sinkmass}
\end{figure*}

This is a particularly interesting possibility for the lowest-mass star, since it may experience additional close encounters as it orbits through and around the stellar system.  If such a star remains below the `survival threshold' of 0.8 M$_{\odot}$ (Figure \ref{sinkmass}), it may be observable in the present-day as a primordial AGB-enriched star in the Milky Way halo or nearby dwarf galaxy.
This smallest sink 
experiences close encounters with the largest sink at, e.g.,
1500 and 2500 yr (red line in right panel of Figure \ref{sinkmass}) that nearly eject it from the disk. 
These encounters occur when the sink has grown to 
only 0.25 M$_{\odot}$ and has a velocity of $\sim$ 5 km s$^{-1}$ relative to the disk.  This is not quite sufficient to escape the stellar system, however.
It is uncertain how much more it will grow as it continues its orbit through the accretion disk, but it is still only 0.25 M$_{\odot}$ at the end of the simulation and may remain below 1 M$_{\odot}$ over its main-sequence lifetime (see also e.g., \citealt{johnson&khochfar2011} for further discussion).

We emphasize the speculative nature of the above scenario.  An AGB phase for the larger protostars of our system would not occur until $\sim 10^8$ yr.  It is at this later time that the smallest protostar's orbit would need to come within $\ga 1$ AU for mass transfer to occur.  While its orbit over the first 5000 yr ranges between one and a few hundred AU, it remains uncertain how the orbit will evolve over the much longer AGB timescales, whether the star will undergo an ejection before this time, etc.  However, in our simulation we do still see the basic initial requirements for our scenaro: that a slowly accreting low-mass star is in close orbit around a larger star on track to eventually undergo an AGB phase.

\begin{figure*}
\includegraphics[width=.4\textwidth]{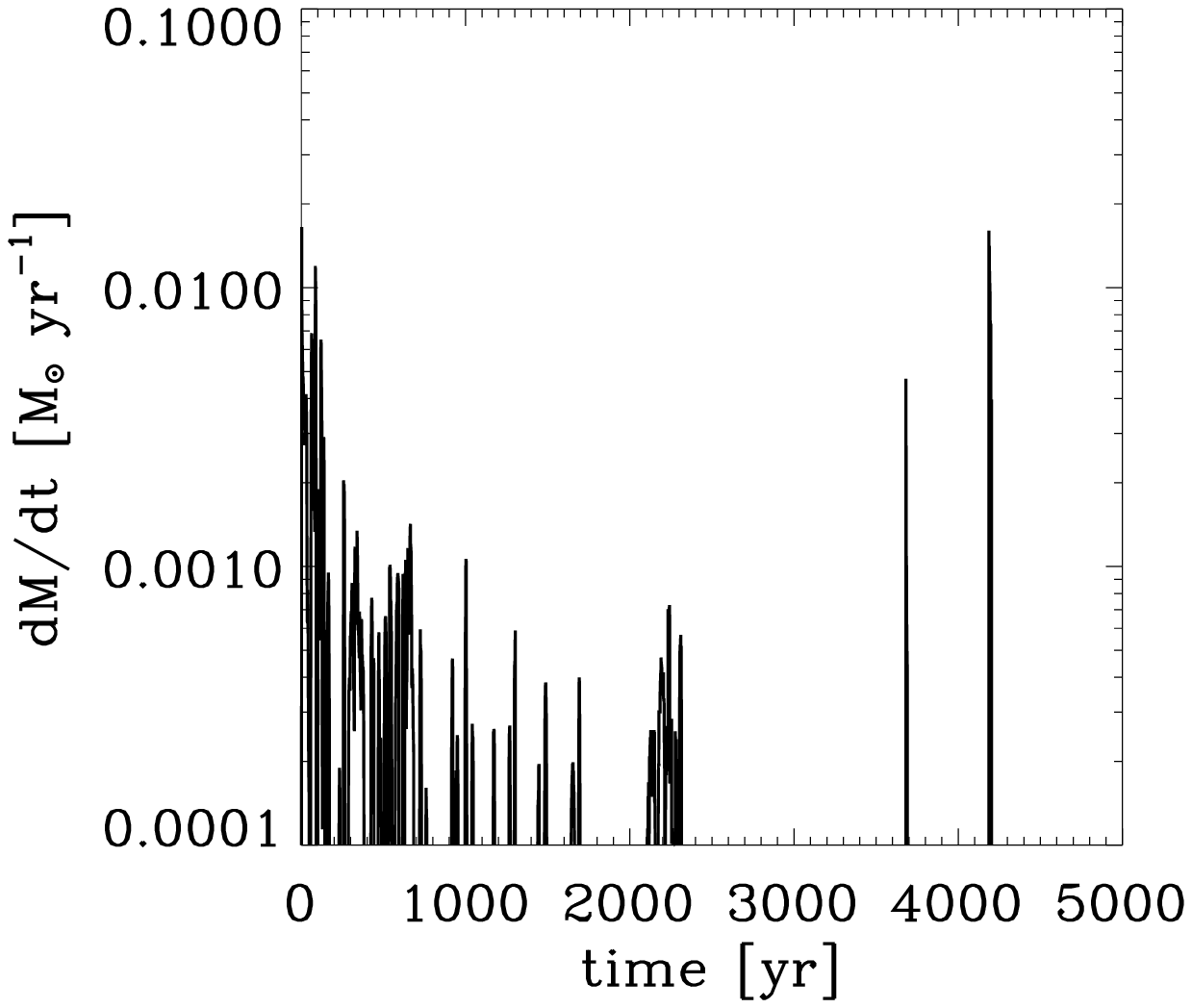}
\includegraphics[width=.4\textwidth]{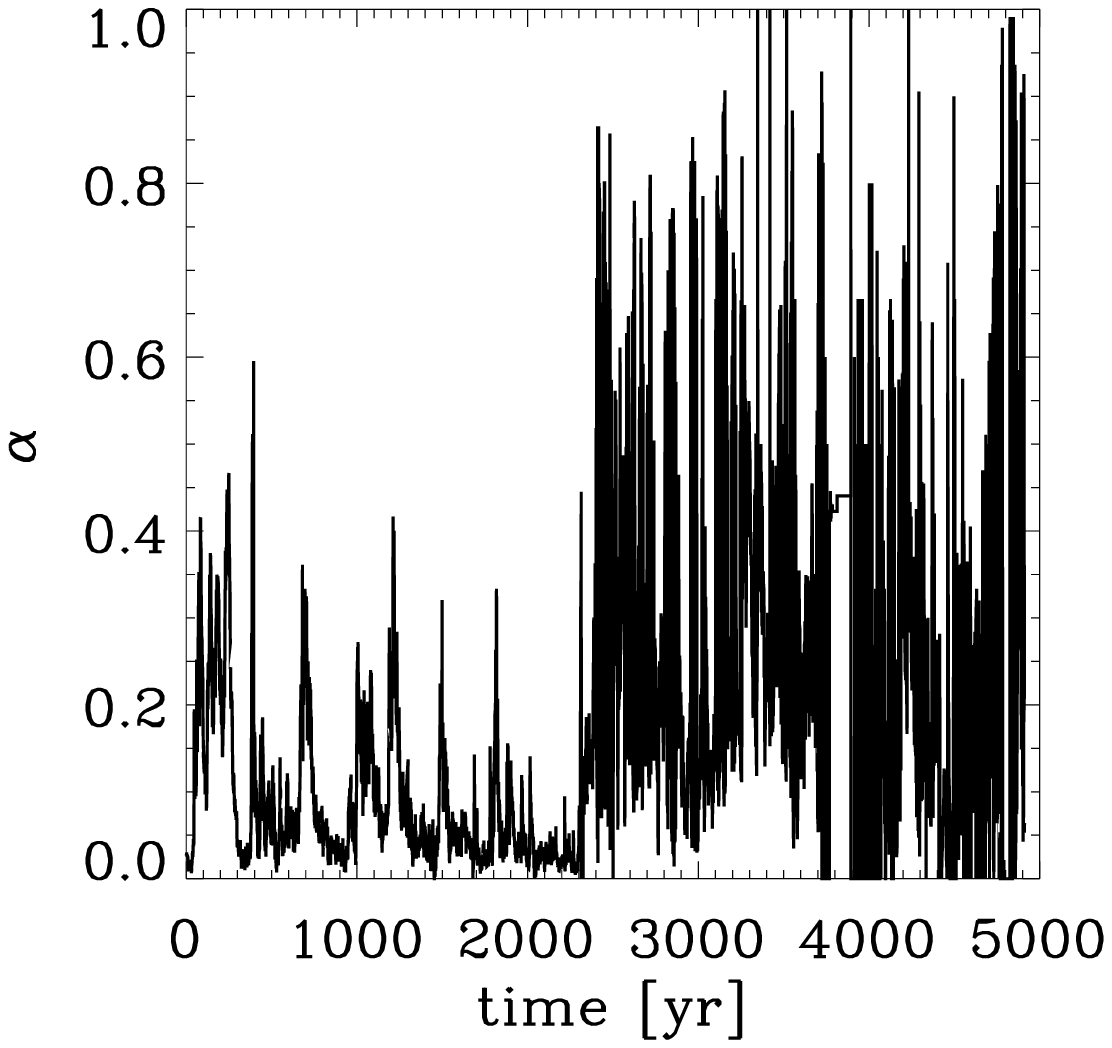}
 \caption{Evolution of various properties of near-sink gas:
 {\it Left:}  Accretion rate onto the main sink. 
{\it Right:} Value of $\alpha$, where $\alpha=1$ corresponds to gas flowing radially towards the sink, while $\alpha=0$ corresponds to a rotationally dominated flow.
%{\it (c):} Ratio of ambient pressure from ionizing radiation $P_{\rm rad}$ to total the total pressure ($P_{\rm tot} = P_{\rm rad} + P_{\rm therm}$).  In the simulation, $P_{\rm rad}$ is included only for particles within the I-front.  Before break-out of the I-front, $P_{\rm rad}$ is calculated for all particles with $n > 10^{13}$ cm$^{-3}$ but not included in the simulation.  
%{\it (d):}  Ratio of radially outward force from ionizing radiation ($F_{\rm rad}$) to inward gravitational force ($F_{\rm grav}$).  $F_{\rm rad}$ is applied only to particles within the I-front.  Before I-front break-out,  $F_{\rm rad}$ is calculated for all particles with $n > 10^{13}$ cm$^{-3}$ for demonstrative purposes, but not included in the simulation.
The sink initially forms from an already disk-like gas configuration, so $\alpha$ is initially very low.  Prior to 2500 yr, phases of more spherical motion around the sink generally correspond to periods of more rapid accretion. 
}
\label{star-model2}
\end{figure*}

The variable orbital motion of the sinks further contributes to the  high variability of the sink accretion rates.  The accretion rate onto the main sink, shown in  Figure \ref{star-model2}, is nearly $\dot{M} \sim 10^{-2}$ M$_{\odot}$ yr$^{-1}$ for the first few hundred years, but quickly drops to $\la$ 10$^{-3}$ M$_{\odot}$ yr$^{-1}$ with periods where $\dot{M} \sim 0$.  As the sink orbits through the stellar disk, the value of $\alpha$ is similarly variable, where $\alpha=1$ corresponds to radially dominated gas motion towards the sink and $\alpha=0$ corresponds to a rotationally dominated flow.  Prior to 2500 yr, periods where $\alpha$ is closer to one corresponds to periods of more rapid accretion.  In the latter half of the simulation when further disk fragmentation occurs, $\alpha$ becomes significantly more variable.  On average, the main sink accretes at $\sim 2 \times 10^{-4}$  M$_{\odot}$ yr$^{-1}$. 
\\
\\
\section{Influence of Lyman-Werner Background}

\subsection{Overview}

A photodisssociating LW background built up by earlier-forming Pop III stars may slow or prevent the cooling and collapse of the gas in our $z=15$ minihalo.  However, the effect of such a background remains very uncertain.   
We first note that our analysis shows it is not simply the redshift but also the rotational structure of the gas that drives the unusually slow infall rate onto this minihalo. There is substantial variation in minihalo characteristics seen at all redshifts.  This implies that such a highly rotationally-supported, slowly accreting gas cloud may also exist within some $z=20-30$ minihalos. 

When considering the effect of LW radiation, it is indeed appropriate to focus on the global background radiation, as opposed to radiation from a particular source.  Given our box size, the nearest minihalo "outside" of the box would be 200 kpc (comoving) away, or $\sim$20 kpc (physical) away.  
%Given the r^-2 dependence of both LW and X-ray flux from a nearby source, the global background should dominate.  
Works by, e.g., \cite{dijkstraetal2008} and \cite{johnsonetal2013} find that it is only rare high-density regions where LW flux from local sources will dominate over the global background, not regions like that in our simulation.  However, the quickness with which this background will grow is very uncertain, and depends upon the early Pop III IMF.  For instance, the semi-analytic models of \cite{crosbyetal2013} find that $J_{21}$ is at least $\sim 1$ at $z=15$, where $J_{21}$ represents units of 10$^{-21}$ erg s$^{-1}$ cm$^{-2}$ Hz$^{-1}$ sr$^{-1}$.  The simulations of \cite{johnsonetal2013}, on the other hand, find at this redshift that the overall $J_{21}$ is an order of magnitude lower, $J_{21}\sim0.1$.  

It is furthermore possible that the concurrently growing X-ray background from Pop III remnant black holes could have an opposing effect to the LW background.  For instance, \cite{jeonetal2012} find that the X-ray radiation from a high-mass BH binary can provide positive feedback such that gas collapse into distant minihalos is facilitated via H$_2$ cooling promoted by the strong X-ray emission.  In contrast, semianalytic models by, e.g., \cite{tanakaetal2012} find that the X-ray heating of the IGM is a stronger effect than the associated enhancement of H$_2$ formation in potential star-forming regions.  

Simulations such as those in \cite{machaceketal2003} and \cite{oshea&norman2008}, which found that the LW background delays gas collapse, did not account for the effect of H$_2$ self-shielding.  They furthermore assumed a $J_{21}$ that was constant instead of gradually evolving.  To properly include the effects of this background, we would need to understand the still uncertain rate at which these backgrounds build up, and we also would need to account for H$_2$ self-shielding within the minihalo.  

\begin{figure}
\includegraphics[width=.4\textwidth]{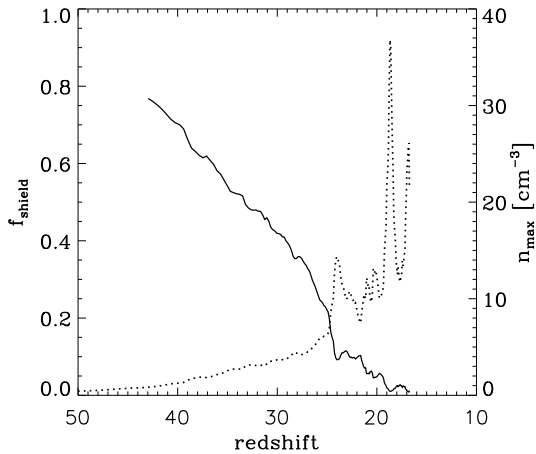}
 \caption{Evolution of $f_{\rm shield}$ with redshift within the gas core of the minihalo.  Dotted line also shows the maximum gas density $n_{\rm max}$.
}
\label{fshield}
\end{figure}

The value of $f_{\rm shield}$ represents the factor by which H$_2$ absorption from the IGM and the outer parts of the minihalo will reduce the local LW flux within the inner star-forming parts of the minihalo. 
Figure \ref{fshield} shows estimates of the average $f_{\rm shield}$ within the gas core of the minihalo, defined as the central gas with densities within a factor of ten of the maximum gas density $n_{\rm max}$.  We estimate $f_{\rm shield}$ in the same manner as in \cite{johnsonetal2013}, in turn based upon \cite{draine&bertoldi1996} and \cite{wolcottetal2011}:

\begin{eqnarray}
f_{\rm shield}(N_{\rm H2}, T) & = & \frac{0.965}{(1+x/b_{\rm 5})^{1.1}} + \frac{0.035}{(1+x)^{0.5}}  \nonumber \\
                   & \times & {\rm exp}\left[-8.5 \times 10^{-4} (1+x)^{0.5} \right] \mbox{\ ,} 
\end{eqnarray}

\noindent where $x$ $\equiv$ $N_{\rm H2}$/5$\times$10$^{14}$ cm$^{-2}$, $b_{\rm 5}$ $\equiv$ $b$/10$^{5}$ cm s$^{-1}$, and $b$ is the Doppler broadening parameter given by $b$ $\equiv$ ($k_{\rm B}$$T$/$m_{\rm H}$)$^{\frac{1}{2}}$ (see \citealt{johnsonetal2013} for further details).  By $z\la25$, $f_{\rm shield}$ will reduce the local LW flux by over an order of magnitude, greatly helping to reduce any possible effects of the LW background.

We finally note that we do not argue that Pop III stars were typically low mass, but that in rare environments, such a low-mass formation mode could occur. This is potentially important as it would allow observers to detect such Pop III fossils as surviving stars in our local cosmic neighborhood (stellar archaeology). The standard picture to date has been that Pop III stars always grow to masses that would have led to their death a long time ago.  Because the strength of the LW background is subject to huge uncertainties, including spatial fluctuations and local opacity, conditions such as those simulated here cannot be excluded.

\subsection{Numerical Tests}

We  further numerically examine LW effects with a set of re-simulations of the initial minihalo collapse, beginning from $z=50$, but this time including a LW background.  We follow the gas collapse up to the point just before sink formation ($n = 10^{16}$ cm$^{-3}$), but we did not have sufficient computational resources to follow the gas evolution further.   The LW background grows with time as

\begin{equation}
J_{21} = J_{21,0} \times 10^{-(z-z_0)/5} \mbox{,}
\end{equation}

\noindent where $J_{21,0}$ is a normalization parameter set to range from 0.1 to 10, and $z_0  = 10$.  When $J_{21,0}=1$, we obtain a good fit to the LW background evolution presented in Figure 1 of \nocite{greif&bromm2006} Greif and Bromm (2006; see also \citealt{pawliketal2013}).  From this we apply a photo-dissociation rate of

\begin{equation}
k_{{\rm H}_{2}} =  1.38\times 10^{-12}\,f_{\rm shield}\,J_{21}~{\rm s}^{-1} \mbox{.}
\end{equation}

We find that the LW background initially serves to delay minihalo collapse.  
For the more extreme $J_{21,0}=10$ case, the minihalo gas still has not collapsed at $z=13.3$.  At this time the densest gas has densities of only $n = 10$ cm$^{-3}$.  To save computational time we did not follow this simulation further.  However, we conclude that at such high $J_{21,0}$ values the gas evolution is significantly altered.  Gas collapse will likely occur only once the minihalo reaches higher mass and virial temperatures of $\sim$ 10$^4$ K.  Gas fragmentation may be suppressed, and the gas may collapse directly into a black hole (see e.g., \citealt{oh&haiman2002, bromm&loeb2003}).

The gas evolution is much less affected for the more physically realistic value of $J_{21,0}=0.1$, in which case collapse to sink particle densities of $n = 10^{16}$ cm$^{-3}$ is delayed by $1.7 \times 10^6$ yr. 
In Fig. \ref{testLW} we show the state of the gas at the point that the gas has reached $n = 10^{16}$ cm$^{-3}$ in each test simulation.  For $J_{21,0} = 0.1$ (solid black line in Fig. \ref{testLW}),  the gas profile does not significantly differ from the fiducial $J_{21,0} = 0$ case (red line in Fig. \ref{testLW}).  As expected, outside of $10^5$ AU the H$_2$ fraction is reduced due to the LW background, while shielding is effective in the more central regions and even allows for a slightly larger H$_2$ fraction inside of 10$^5$ AU. In the inner 1000 AU, the gas properties do not differ by more than a few tens of percent, though the central densities and estimated spherical accretion rate (see Equ. 14) are somewhat enhanced.  
The enclosed mass at all given radii,  as well as the ratio of $M_{\rm enc}$ to $M_{\rm BE}$, are also slightly larger (Fig. \ref{menc_testLW}).
This leads to a slightly higher overall disk mass as the gas approaches $n = 10^{16}$ cm$^{-3}$, when $M_{\rm disk} \sim 20$ M$_{\odot}$  as opposed to $\sim 16$ M$_{\odot}$ in the fiducial case (Fig. \ref{diskmass_testLW}).  Note, however, that these values of $M_{\rm disk}$ and $\dot{M}_{\rm sphere}$ are approximately half to a tenth as large as those found in the other minihalos discussed in Section 4.2.  
Thus, under a small $J_{21,0} = 0.1$ background we would still likely find an unusually low-mass Pop III system, though probably more massive than the fiducial case.

 For $J_{21,0}=1$, the differences are more significant (dashed line in Fig. \ref{testLW}).  Collapse is delayed by $3.3 \times 10^7$ yr before the gas finally reaches the sink density of $n = 10^{16}$ cm$^{-3}$.  The H$_2$ fraction is approximately half of that found in the fiducial case at distances greater than 10$^4$ AU.  However, inside 10$^4$ AU shielding becomes very effective, and the H$_2$ increases more rapidly than the other test cases as the radius declines.  In the central 10$^4$ AU the gas density, sound speed, and infall velocity are all reduced, and $\dot{M}_{\rm sphere}$ is up to an order of magnitude below the fiducial case.  
 The enclosed mass at all given radii is also reduced by up to a factor of a few (Fig. \ref{menc_testLW}).  
 The central 30 $M_{\rm enc}$ has a higher ratio of $M_{\rm enc}$ to $M_{\rm BE}$, but outside of this region the ratio is much reduced and the gas is more gravitationally stable.
 Overall, this leads to $M_{\rm disk} \sim 5$ M$_{\odot}$, when $n = 10^{16}$ cm$^{-3}$, several times smaller than the other cases. 
 
When comparing these simulations at the same peak densities, but not at the same physical time, the effect of increasing the LW background is not monotonic.  For the more realistic values of $J_{21,0} = 0.1$ or 1, we find central accretion rates that are either slightly enhanced or significantly reduced.  In these cases the LW background does not seem to change our general finding of a minihalo which hosts a system of very slowly accreting Pop III stars.

\begin{figure*}
\includegraphics[width=.4\textwidth]{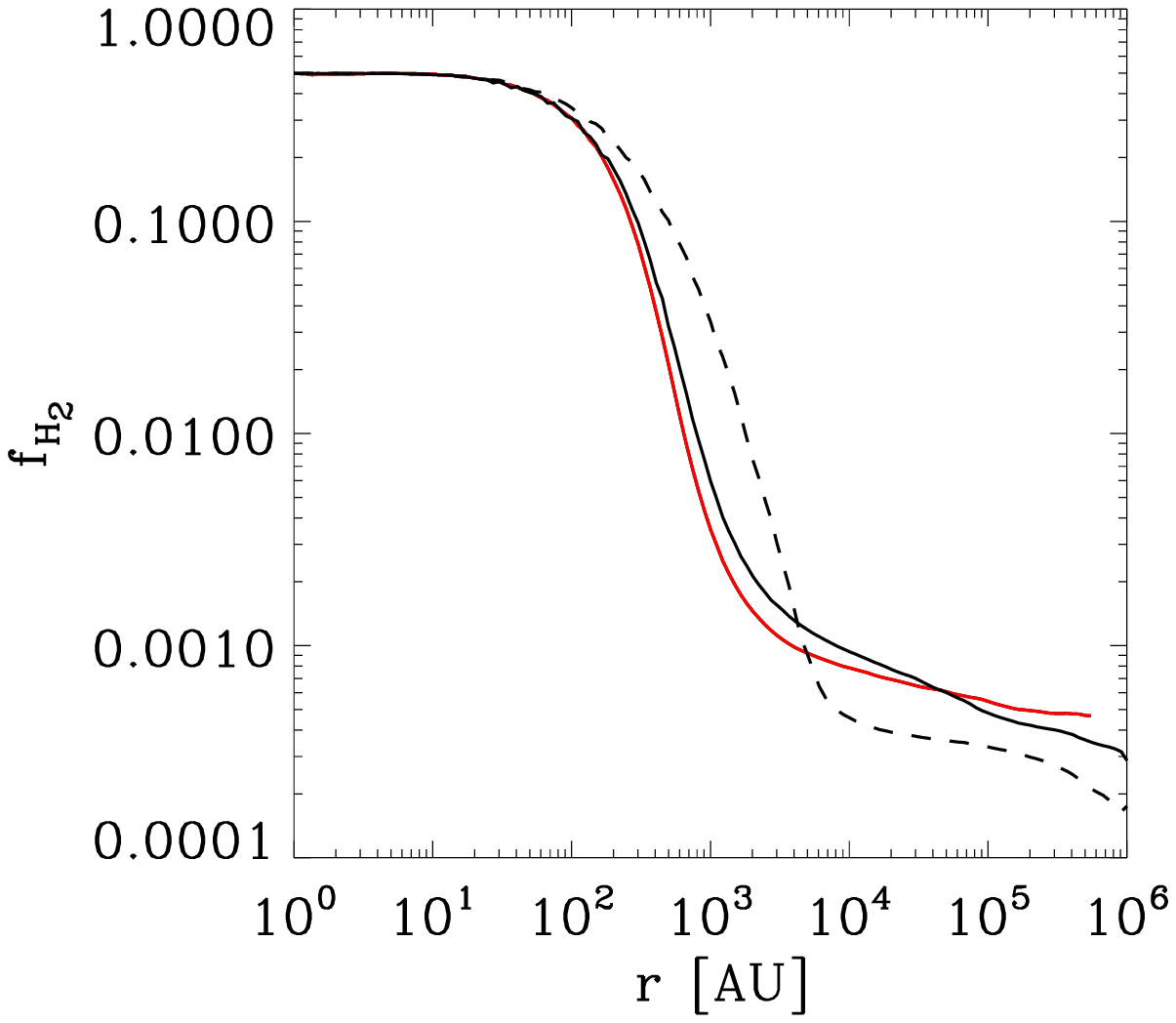}
\includegraphics[width=.4\textwidth]{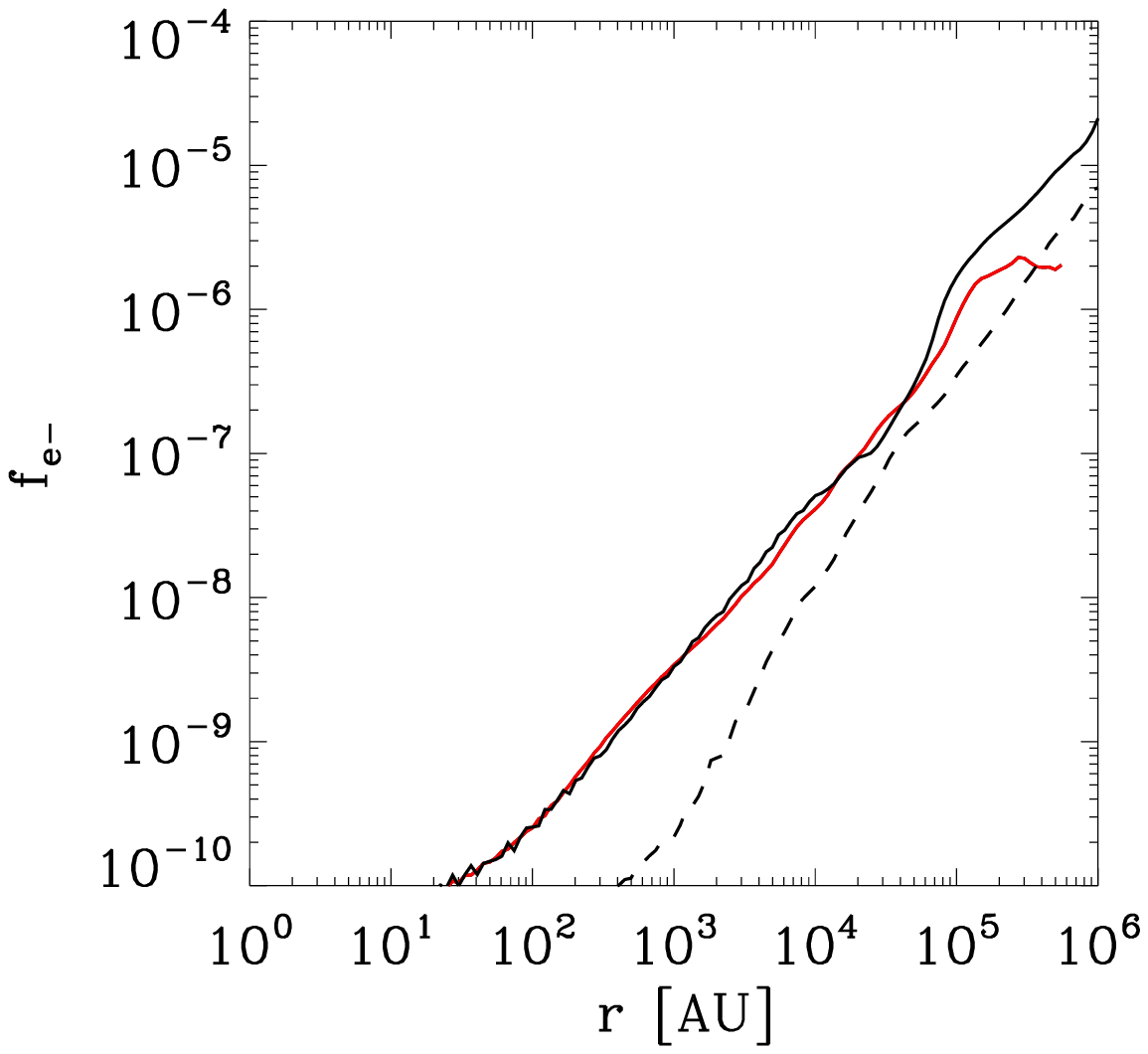}
\end{figure*}
\begin{figure*}
\includegraphics[width=.4\textwidth]{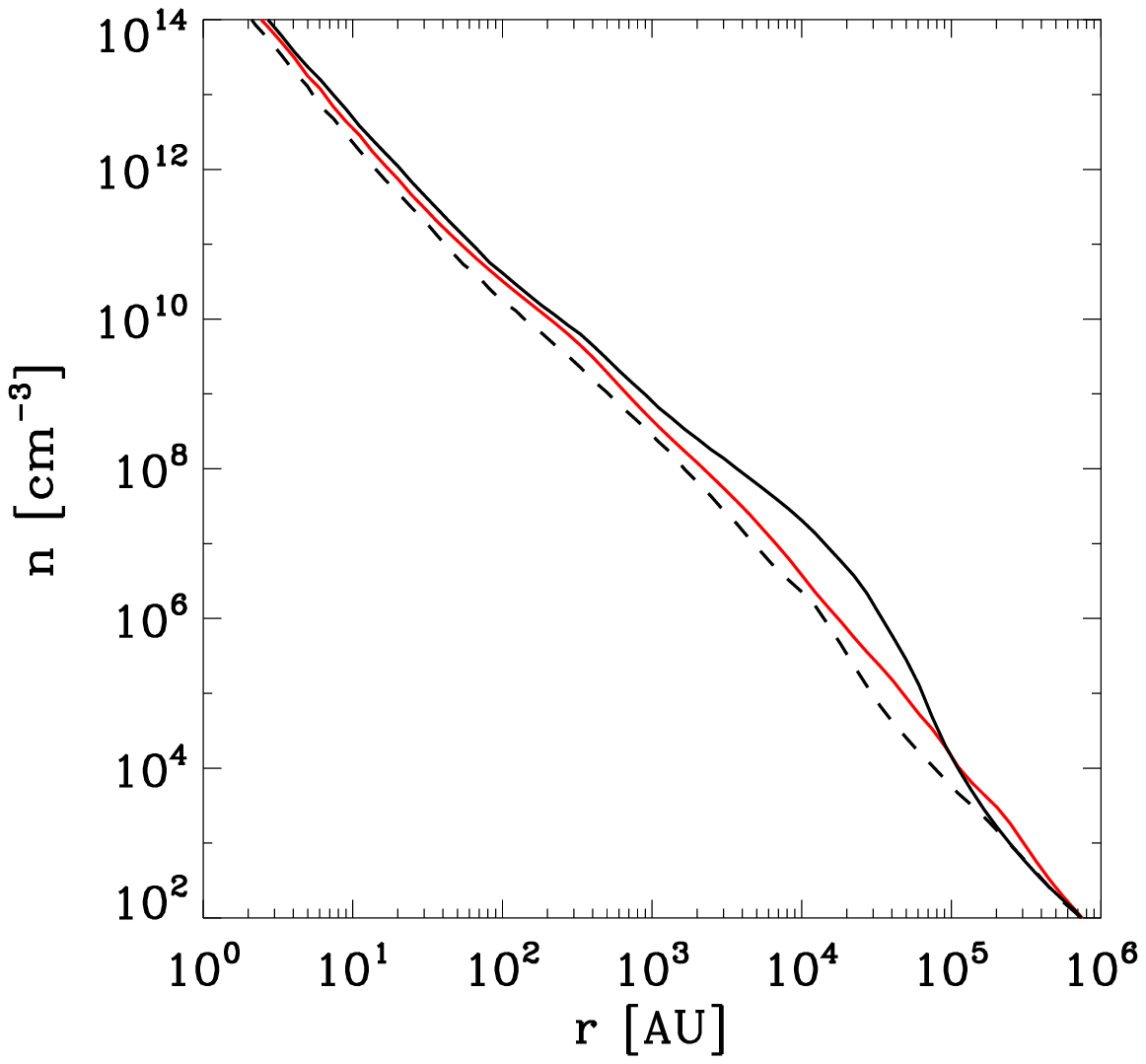}
\includegraphics[width=.4\textwidth]{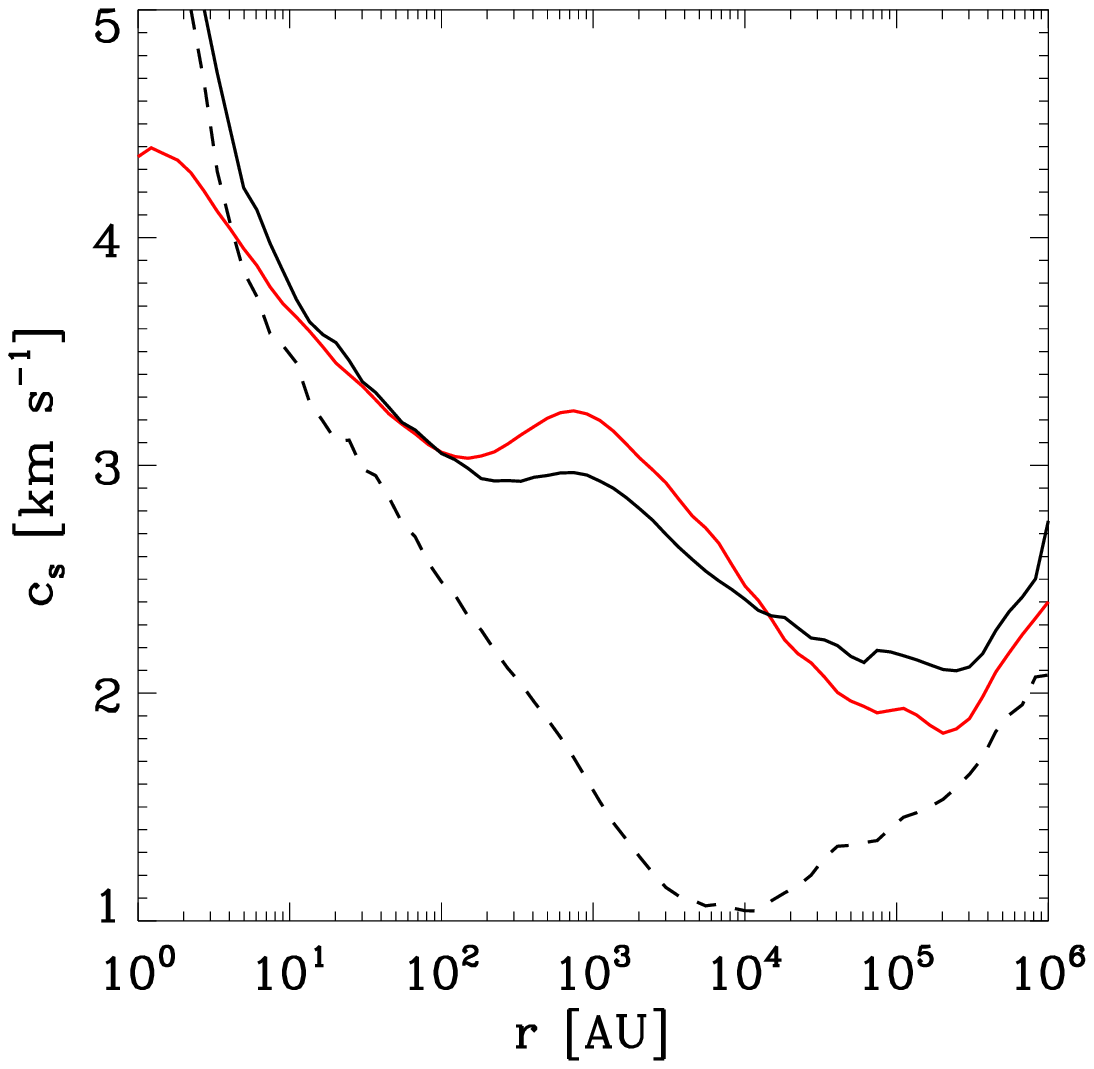}
\end{figure*}
\begin{figure*}
\includegraphics[width=.4\textwidth]{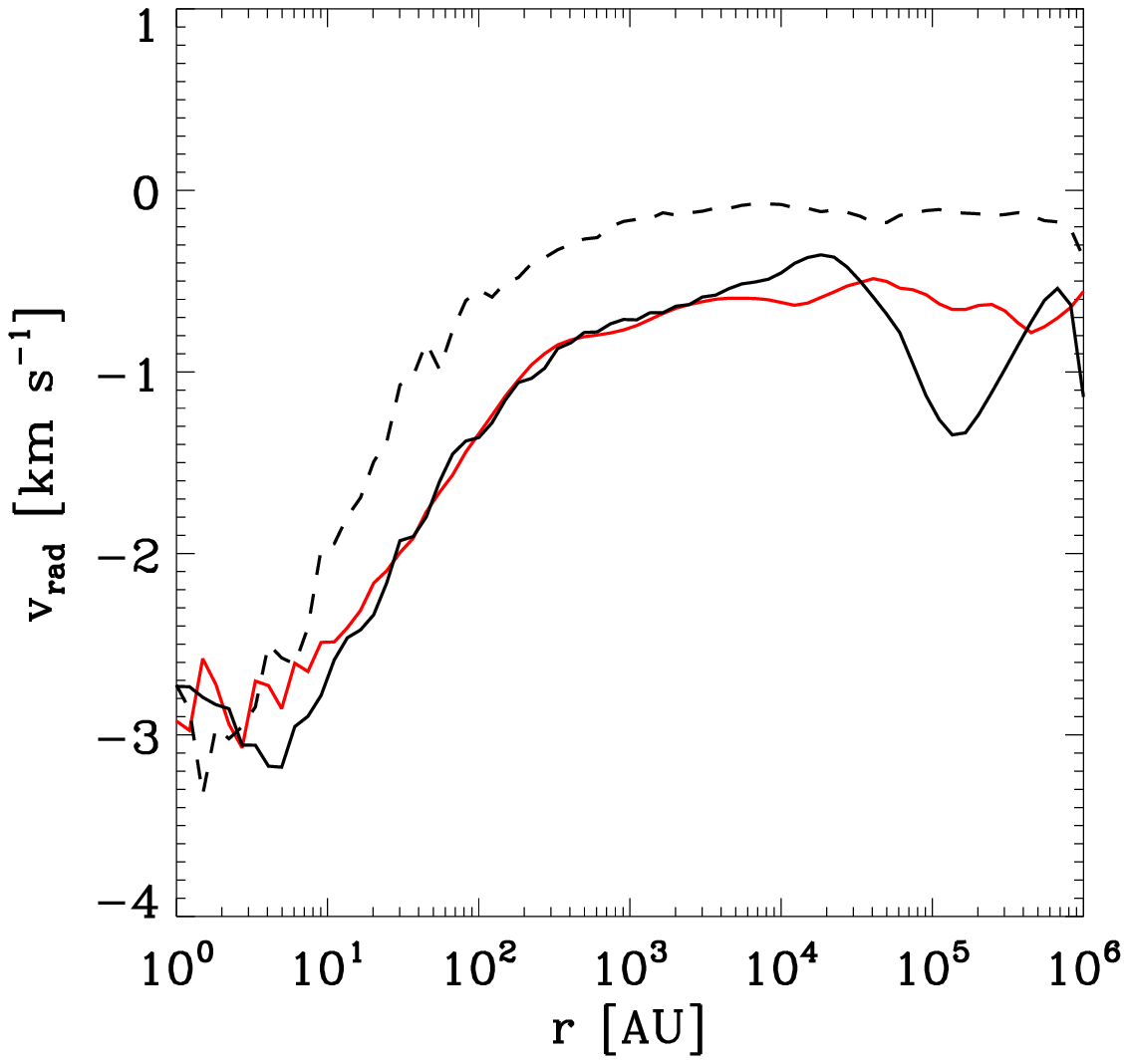}
\includegraphics[width=.4\textwidth]{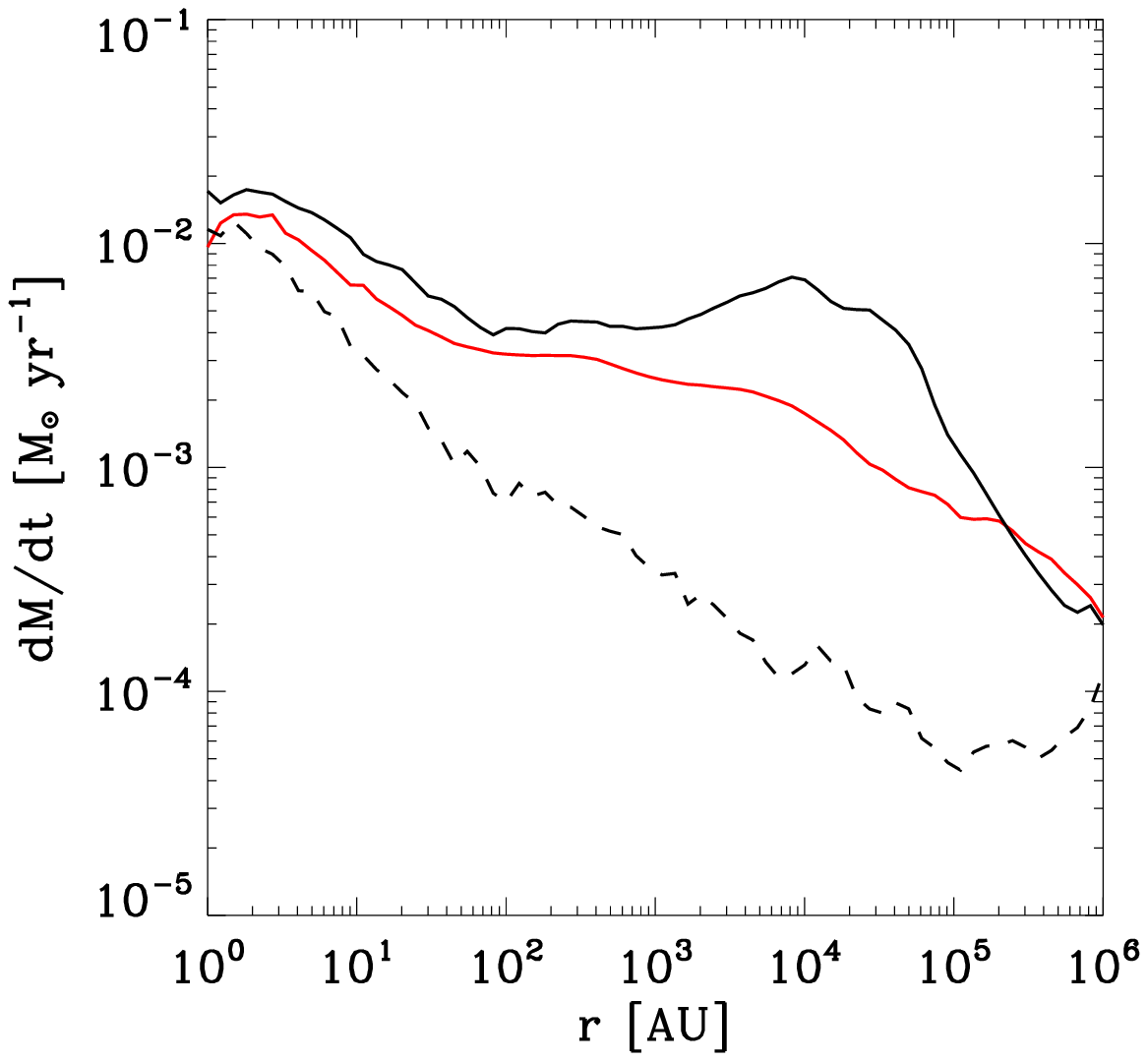}
 \caption
 {
Radial profiles of the central gas within our simulated minihalo under various strengths of LW background, measured when the gas has reached a density of $n = 10^{16}$ cm$^{-3}$ in the respective simulations.  Solid black lines denote the $J_{21,0} = 0.1$ case, while dashed lines represent  $J_{21,0} = 1$. 
%and dotted lines denote  $J_{21,0} = 10.0$.  
Red lines are taken from the fiducial $J_{21,0} = 0$ simulation discussed throughout this work.
{\it Top Left:} H$_2$ fraction.
{\it Top Right:} Electron fraction.
{\it Middle Left:} Number density.
{\it Middle Right:} Sound speed.
{\it Bottom Left:} Radial infall velocity.
{\it Bottom Right:} Estimated spherical accretion rate ($\dot{M}_{\rm sphere}$, see Equ. 14).
The gas profiles are not significantly altered for $J_{21,0} = 0.1$, though density, infall velocity, and $\dot{M}_{\rm sphere}$ are slightly enhanced.  
The $J_{21,0} = 1$ background surprisingly leads to the generally opposite effect of reduced densities, sound speeds, and infall rates.
 }
\label{testLW}
\end{figure*}

\begin{figure}
\includegraphics[width=.45\textwidth]{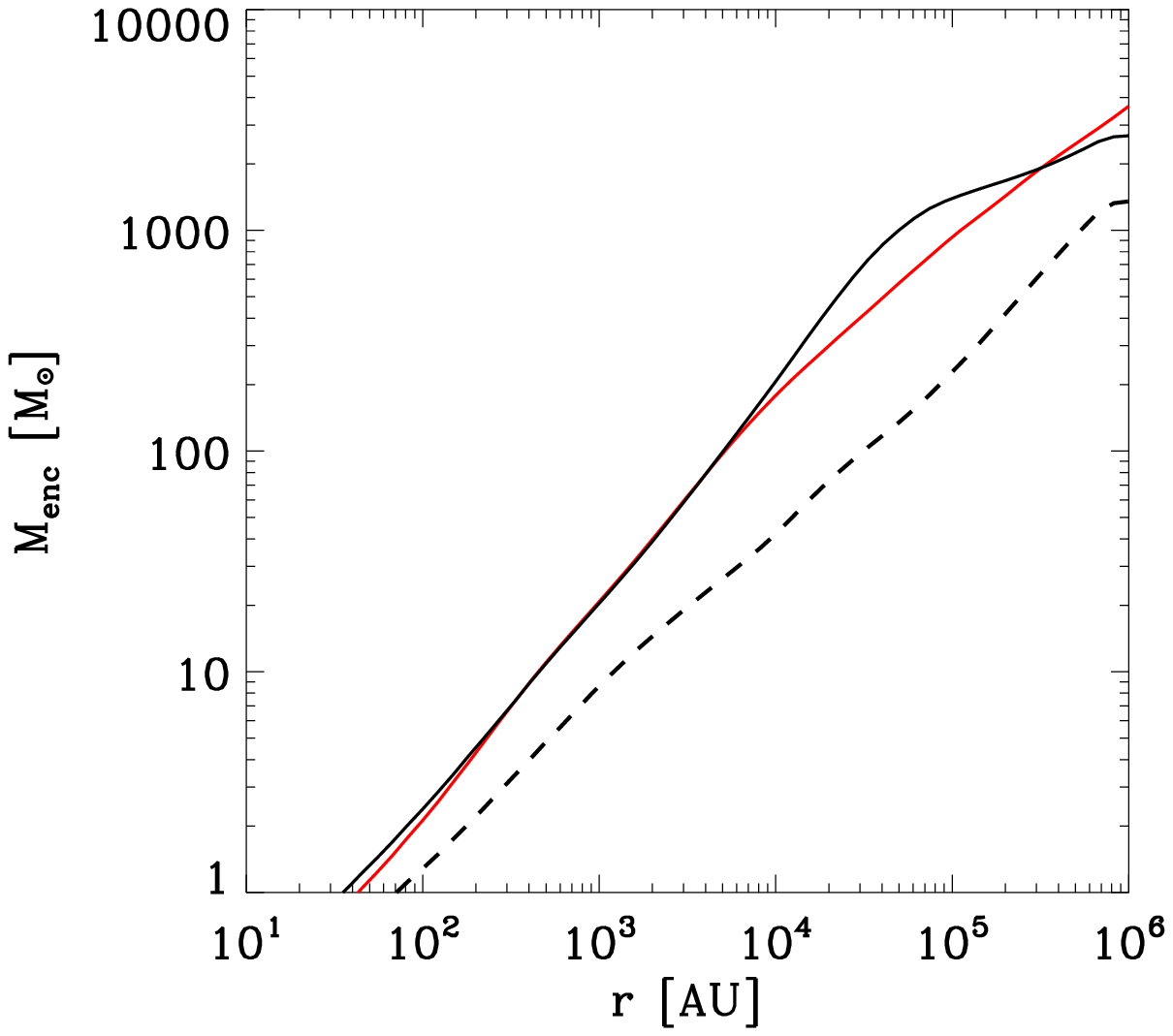}
\includegraphics[width=.45\textwidth]{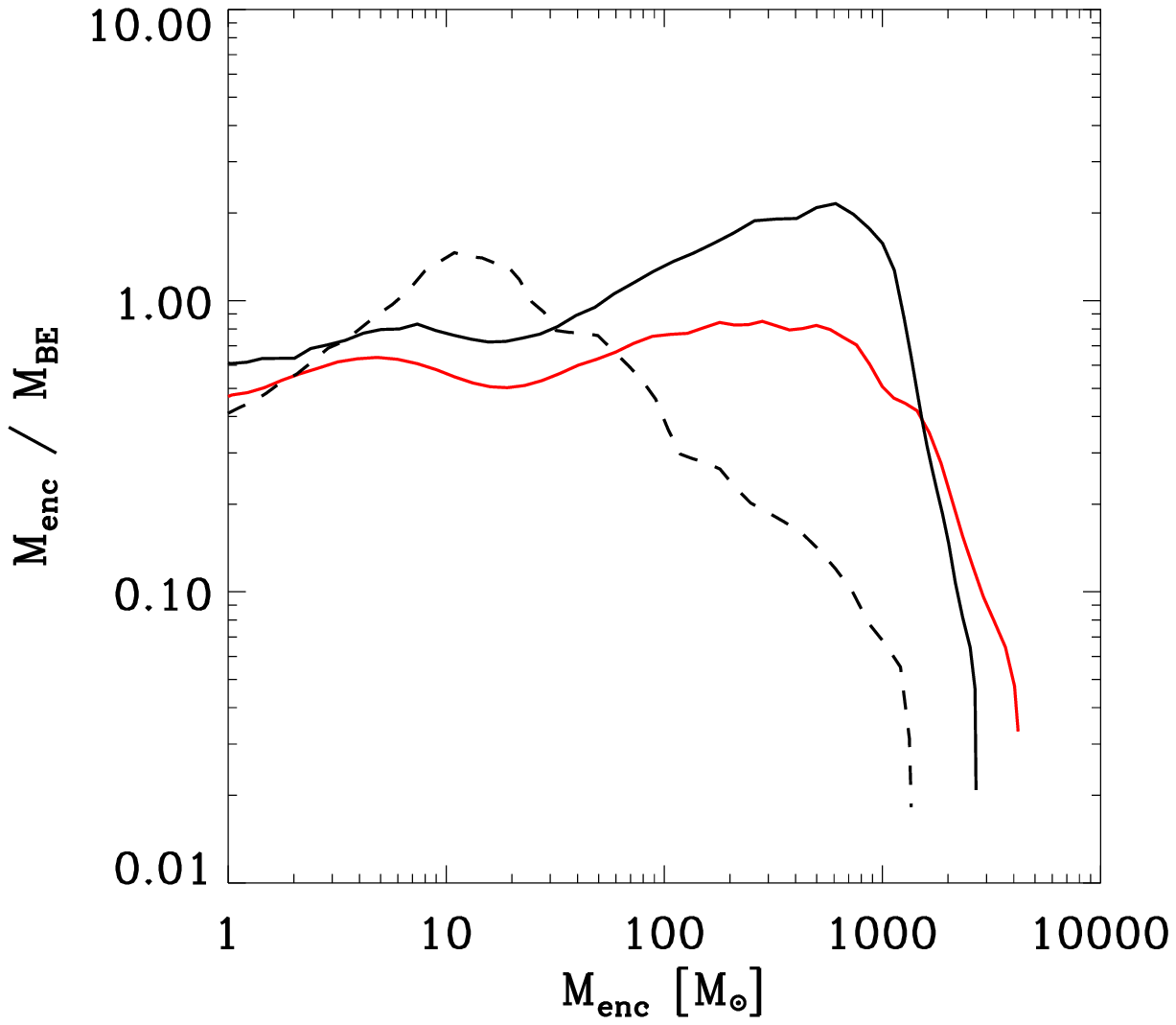}
 \caption
 {
 {\it Top:} Enclosed mass $M_{\rm enc}$ versus radius for our test cases at the point when the maximum gas density first reaches $10^{16}$ cm$^{-3}$.
  Solid black lines denote the $J_{21,0} = 0.1$ case, while dashed lines represent  $J_{21,0} = 1$. 
Red lines are taken from the fiducial $J_{21,0} = 0$ simulation discussed throughout this work.
The central point is taken to be the most dense gas particle.
{\it Bottom:} Ratio of  $M_{\rm enc}$  to $M_{\rm BE}$ versus  $M_{\rm enc}$.   Lines have same meaning as in the upper panel.
While the $J_{21,0} = 0.1$ case has slightly enhanced mass and $M_{\rm enc}$  to $M_{\rm BE}$ ratio, these are somewhat reduced for the $J_{21,0} = 1$ case.
  }
 \label{menc_testLW}
\end{figure}

\begin{figure}
\includegraphics[width=.45\textwidth]{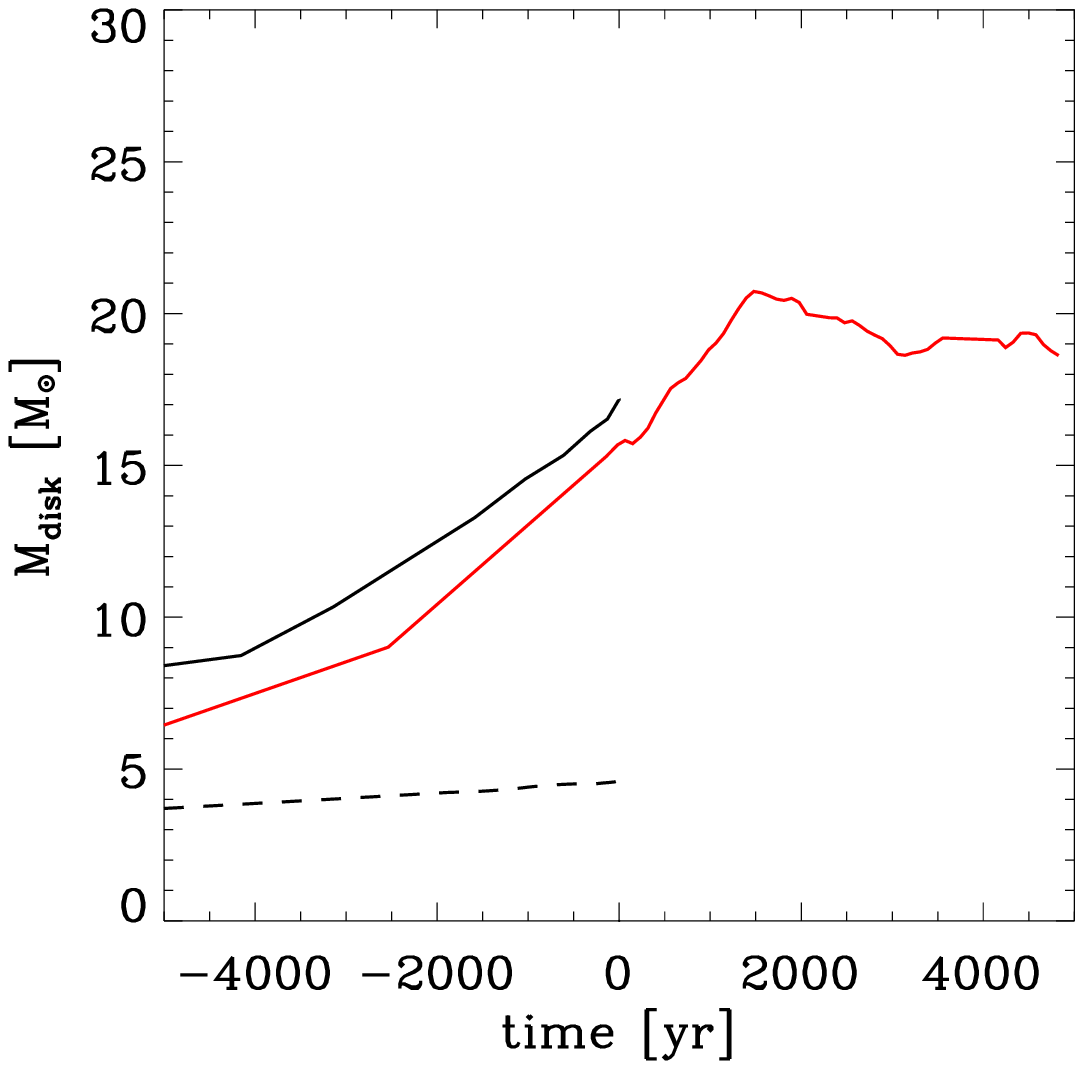}
 \caption
 {
 Evolution of disk mass over time for $J_{21,0} = 0$, $J_{21,0} = 0.1$, and $J_{21,0} = 1$.  Line styles have the same meaning as in the previous figure.  Time is measured with respect to the time when the gas first reaches $n = 10^{16}$ cm$^{-3}$.  The $J_{21,0} = 0.1$ background leads to a slightly enhanced $M_{\rm disk}$ as compared with the fiducial $J_{21,0} = 0$ model.  The larger $J_{21,0} = 1$ background in fact leads to $M_{\rm disk}$ which is consistently several times smaller than the fiducial case.  Realistic LW backgrounds are therefore unlikely to change our overall finding of unusually low Pop III accretion rates within our simulated minihalo.
 }
 \label{diskmass_testLW}
\end{figure}

\section{Discussion and Conclusions}

We present a three-dimensional simulation of the formation and growth of a Pop III stellar system.  This calculation was initialized on cosmological scales while resolving lengths as small as 1 AU. We found that the host minihalo formed at an unusually low redshift of $z=15$, leading to a low DM and baryonic accretion rate as well as a low-mass and slowly-growing stellar system.  The stars in the system can be expected to reach $\sim$ 0.5 to 5 M$_{\odot}$ after 1 Myr.  This is nearly an order of magnitude slower than rates typically found within $z\ga 20$ stellar systems (e.g., \citealt{stacyetal2010,greifetal2011, smithetal2011}).  
We additionally find that a LW background as high as $J_{21,0} = 1$ will not significantly change our finding of uncharacteristically small infall rates onto the central star-forming region of our minihalo. 

It is uncertain how common such low-mass Pop III systems will be.
The minihalo we present here hosts the slowest-accreting Pop III system of approximately 10 minihalos which are included in our comparison in Section 4.  \cite{hiranoetal2014} use approximately 100 minihalos from cosmological simulations to initialize two-dimensional simulations of Pop III stellar growth under feedback.  Only one star can be followed per minihalo, and in this case their smallest star is expected to grow to $\sim$ 9 M$_{\odot}$.  Were they able to follow fragmentation within this particular minihalo, it is possible that the stellar mass may have instead been distributed among several stars of lower mass.  We thus make a very rough estimate that one out of a few tens to one out of 100 minihalos will host a low-mass system similar to what we find in our simulation.  
Considering that the mass of 10$^5$ to 10$^6$ minihalos will ultimately become incorporated into a Milky Way-type galaxy, it is conceivable that on the order of thousands of Pop III stars from such low-mass systems may exist in the nearby Galactic halo.

Further study will be necessary to more precisely determine 
how common such low-mass Pop III systems are and 
whether our $z=15$ Pop III system indicates a more general transition in the Pop III IMF from $z\sim30$ to $\sim10$, or if our particular minihalo was an unusual case even for its low redshift.  While minihalo comparisons of other works do not find this same transition in accretion rate (\citealt{gaoetal2007, oshea&norman2007}), the numerical analyses by \cite{desouzaetal2013} do find a transition in the spin distribution of gas within minihalos as redshift declines.  According to the semi-analytic model of \cite{mckee&tan2008}, such differences in rotational support will in turn lead to differences in the protostellar feedback and the final mass reached by the protostar.   \cite{desouzaetal2013} argue that this will cause the Pop III IMF peak to shift to lower mass with lower redshift.  This remains to be tested with more physically detailed simulations.

Understanding the Pop III IMF evolution with redshift will additionally require a greater knowledge of how the build-up of global background radiation as well as magnetic fields proceeded at this redshift (e.g., \citealt{schleicheretal2010, schoberetal2012, turketal2012}). 
Understanding how the IMF evolved to later times is particularly important given that recent work has indicated that metal-free gas will indeed survive to relatively low redshift.  For instance, \cite{simcoeetal2012} reported observations of extremely low metallicity or possibly metal-free gas within a $z\sim7$ damped Ly-$\alpha$ system, while \cite{fumagallietal2011} reported the detection of metal-free gas within Lyman-limit systems at $z\ga3$.
Numerical work (e.g., \citealt{muratovetal2013}, see also \citealt{scannapiecoetal2005}) has similarly indicated that Pop III star formation can continue to $z\sim6$.
Our preliminary tests presented in Section 5, however, indicate that for a range of LW backgrounds our Pop III system will still undergo unusually low accretion rates.

AGB stars are important to understanding the early chemical evolution of the galaxy
(e.g., \citealt{karakas2010,campbell&lattanzio2008,karakas2010, karakas&lugaro2010}). 
They are known to produce significant quantities of carbon, nitrogen, and s-process elements. 
(\citealt{bussoetal2001, siessetal2002, siess&goriely2003}).
They may also help to explain the observation of large amounts of dust in high-$z$ galaxies and quasars,
 which implies rapid dust production in the early universe 
(\citealt{bertoldietal2003,valianteetal2009,galletal2011}).  
Our results reveal a pathway for such AGB star formation within primordial gas.
 In particular, the larger 3-5 M$_{\odot}$ stars of our simulation are sufficiently short-lived to undergo a metal and dust-enriching AGB phase by $z>6$, while the smaller $\la$ 1 M$_{\odot}$ star is long-lived enough that it may still be observed as carrying the enrichment signatures of its larger companions.
The physical scenario suggested by our simulations may indeed explain the abundances of 
certain metal-poor stars in the Milky Way halo, particularly those with unusual features in C, N, and O.  
Studies have found that some of these may be Pop III stars which received material from an AGB companion that is now a white dwarf (e.g., \citealt{sudaetal2004, sudaetal2013}).  
% they eventually did find a definite detection for Ca, albeit at extremely low levels. This Ca cannot easily be explained with AGB models; thus, they now interpret the pattern as coming from a ~60 M_sun faint SN. We should still work this into our conclusion, mentioning this SkyMapper star, which demonstrates its potential to detect a possible AGB signature (just CNO, and zero iron peak and alpha, including Ca). "It is thus an empirically testable question whether an exotic Pop III low-mass mode existed." Something like this. Maybe, we can touch base briefly to fine-tune things.

Studies over recent years have found increasing complexity in the nature of Pop III stars.  Though dust and metallicity by definition do not play a role in primordial star formation, other physical processes such as 
multiplicity (e.g., \citealt{clarketal2008, stacyetal2010, greifetal2011}) as well as the binary nature and rotation rate of Pop III stars (e.g., \citealt{stacy&bromm2013}) will also be of crucial importance.  
In addition, feedback (e.g., \citealt{hosokawaetal2011, smithetal2011, stacyetal2012}) and magnetic fields play a central role in Pop III protostellar accretion.  
However, our work shows that even when such processes are not included in simulations, the differences in minihalo environments alone can lead to substantial variation between primordial stellar clusters.  
As shown by \nocite{jappsenetal2009} Jappsen et al. (2009, see also \citealt{dopckeetal2013}), our work demonstrates that the transition to a low-mass IMF will depend upon not only a critical metallicity (e.g., \citealt{brommetal2001}), but also upon the characteristics of the primordial star-forming region.  

We thus predict the rare existence of low-mass Pop III stars that have survived until the present day, and that may show evidence of enrichment from a companion's AGB-phase mass overflow.
Continued improvements in simulation techniques and computational power, as well as constraints provided by observations such as abundance measurements of metal-poor stars in the Galactic halo and dwarf galaxies (e.g., \citealt{beers&christlieb2005,frebeletal2005,caffauetal2011}; see also \citealt{karlssonetal2013}), 
will provide an increasingly refined picture of the role Pop III stars played in shaping the early universe.
Ultimately, the question of whether
   true Pop~III survivors exist, e.g., in the guise of AGB self-enrichment as
   suggested in this paper, is a question that can be tested empirically. The
   potential for such stellar archaeological constraints is demonstrated by
   the exciting recent discovery by the SkyMapper Southern Sky Survey of a
   star with no detected Fe-peak elements, but low abundances of C, N, O
   (\citealt{kelleretal2014}). Indeed, the detected abundance pattern closely
   resembles an AGB self-enrichment pattern with the one crucial exception
   of an extremely low, but non-zero, Ca abundance. The latter cannot be accommodated
   with AGB enrichment, but instead points to the signature of supernova enrichment from
   a massive Pop~III progenitor, rendering this extreme star second-generation.
   However, the search for Pop~III fossils is clearly within the reach of current
   and upcoming surveys.

\section*{Acknowledgements}

The authors thank the anonymous referee which helped us to improve this manuscript.
AS is grateful for support from the JWST Postdoctoral Fellowship through the NASA Postdoctoral Program (NPP).  
VB acknowledges support from
NASA through Astrophysics Theory and Fundamental
Physics Program grant NNX09AJ33G and from NSF through
grant AST-1009928.
Resources supporting this work were provided by the NASA High-End Computing (HEC) Program through the NASA Advanced Supercomputing (NAS) Division at Ames Research Center.

\bibliographystyle{apj}
\bibliography{agb}

\clearpage

\end{document}